\documentclass[12pt]{article}
\usepackage{graphicx}
\usepackage{mathrsfs}
\usepackage{dsfont}
\usepackage{graphicx}
\usepackage{units}
\usepackage{amsmath, amsthm, amssymb, amsfonts, enumerate}
\usepackage{natbib}
\usepackage{color}
\usepackage{caption}
\usepackage{subcaption}
\usepackage{booktabs}
\usepackage{setspace}
\usepackage{mdframed}
\usepackage{hyperref}
\usepackage{float}
\usepackage{multirow} 
\usepackage[normalem]{ulem}
\usepackage{placeins}
\usepackage{tikz}
\usepackage{pgfplots}
\pgfplotsset{compat=newest}
\usetikzlibrary{calc}
\usepackage{soul}

\usetikzlibrary{calc}
\usepackage[left]{lineno}

\definecolor{Red}{rgb}{1, 0.0, 0}
\definecolor{blue}{rgb}{0.0, 0.0, 1}
\definecolor{green2}{rgb}{0.0, 0.52, 0.24}
\definecolor{cadmiumgreen}{rgb}{0.0, 0.42, 0.24}
\definecolor{camouflagegreen}{rgb}{0.47, 0.53, 0.42}
\definecolor{darkolivegreen}{rgb}{0.33, 0.42, 0.18}
\definecolor{darkpastelgreen}{rgb}{0.01, 0.75, 0.24}
\definecolor{darkspringgreen}{rgb}{0.09, 0.45, 0.27}
\definecolor{darkspringgreen}{rgb}{0.09, 0.45, 0.27}
\usepackage{comment} 

\newcommand{\E}{\mathbb{E}}

\newcommand{\heading}[1]{\vskip .3cm \noindent \textbf {#1} \hskip .4cm }

\def\O{\Omega}
\def\o{{\omega}}
\def\G{\Gamma}

\def\f{\varphi}

\long\def\ignore#1{}

\def\1{\emph{\textbf{1}}}

\usepackage[letterpaper,margin=1.25in]{geometry}
\usepackage{setspace}
\newtheorem{theorem}{Theorem}

\newtheorem*{theorem*}{Theorem}
\newtheorem{proposition}{Proposition}
\newtheorem{corollary}{Corollary}
\newtheorem{example}{Example}
\newtheorem{definition}{Definition}

\newtheorem{remark}{Remark}
\newtheorem{lemma}{Lemma}
\newtheorem{obs}{Observation}

\begin{document}

\title{The Structure of Joint Posterior Beliefs\thanks{For their valuable comments, the authors wish to thank participants of the $37^{\rm th}$ Stony Brook Game Theory Conference, the Paris Workshop on Game Theory and Language: Half a Century of ``Agreeing to Disagree'', the University of Toronto Economic Theory Workshop, the Durham University Economics Seminar, the Adam Smith Business School Micro Theory Seminar of Glasgow University, INSEAD EPS seminar, The School of Economics seminar at the University of Edinburgh, Western University Economics seminar, the Rationality Center Game Theory Seminar, the Technion Game Theory Seminar, the Bar-Ilan University Theoretical Economics Seminar, the Tel Aviv University Game Theory Seminar,  the Bar-Ilan University Management Seminar and the BGU Economics seminar. Lagziel acknowledges the support of the Israel Science Foundation, Grant \#2074/23. Lehrer acknowledges the support of the Deutsche Forschungsgemeinschaft (DFG, German Research Foundation) -- Project Number 461570745. }}
\author{David Lagziel\thanks{Department of Economics, Ben-Gurion University of the Negev, Beer-Sheba 8410501, Israel.  E-mail: \textsf{davidlag@bgu.ac.il}. Visiting Professor, Department of Economics, University of Toronto, 2025–2027.} 
\and
Ehud Lehrer\thanks{Economics Department, Durham University, Durham DH1 3LB, UK.  E-mail: \textsf{ehud.m.lehrer@durham.ac.uk}.}
}
\date{\today}
\maketitle

\thispagestyle{empty}

\begin{abstract}
\singlespacing{
We analyze information structures in environments with privately informed agents and a partially informed mediator who provides public signals. Necessary and sufficient conditions, termed \textit{internal} and \textit{external consistency}, are derived for the rationalizability of joint posteriors. Analyzing the geometry of the feasible set of posteriors shows that informational constraints alone can generate non-convexity.  We also characterize the convexity of the feasible set in the single-agent setting. Applications to belief depolarization, information design, Bayesian persuasion, and potential games are discussed.
}
\end{abstract}

\bigskip

\noindent {\emph{Journal of Economic Literature} classification numbers: C72, D82, D83.}

\bigskip


\newpage
\setcounter{page}{1}

\section{Introduction} \label{Section - Intro}

What restrictions does a partially informed mediator face when shaping the beliefs of privately informed agents? A public signal must be based on the mediator's information, yet its interpretation depends on each agent's private information. Even with a single agent, the beliefs induced at different private information cells cannot generally be chosen independently. Our paper studies these joint feasibility restrictions.

We study a setting with privately informed players and a partially informed external information provider (a \emph{mediator}), all defined on a finite state space with a common prior. Each player and the mediator are endowed with a partition of the state space. The mediator sends a \emph{public} signal according to a stochastic rule that is measurable with respect to its information. Upon observing the signal, players update their beliefs by Bayes' rule, generating a \emph{joint posterior profile} across players (and across states at which the signal can occur). Our central question is a feasibility one: given a joint posterior profile, when does there exist a mediator-compatible public signal that induces it? We answer this by providing necessary and sufficient conditions for implementability.

This implementability concern is central in economics and game theory, where Bayesian updating is the cornerstone of belief revision under uncertainty. In strategic environments with incomplete information, players use signals to update their beliefs about the underlying state of the world, often relying on publicly observed information.  While it might be possible to construct posterior beliefs through arbitrary signaling mechanisms, we focus on the more subtle problem of whether such posteriors can be implemented under the restriction that signals must be measurable with respect to the mediator’s limited knowledge.
Moreover, the mediator's partition of the state space may differ from those of the players, creating the novel possibility that players have common knowledge of events that the mediator does not know.

Our main tool is the \emph{graph of information}, following the framework of \cite{Rodrigues-Neto2009}, which encodes the state space and players' informational partitions in a concise form. On this graph, we define a \emph{posterior likelihood function}~$\f$ that records posterior odds divided by prior odds across \emph{adjacent states} (i.e., states lying in the same information set of some player). For an implementable profile, these are the corresponding signal likelihood ratios. We then ask under what conditions such a function can be rationalized by a mediator-generated signal.

While the graph provides a convenient representation of informational constraints, the core of our analysis lies in understanding how these constraints shape the implementability of belief updates. The function~$\f$ is particularly valuable in our set-up because the likelihood ratios it encodes are shared by all players who cannot distinguish between adjacent states.  As a result, the information carried by~$\f$ is adequate for our analysis, rendering the graph-based model both concise and sufficiently informative. 

Our main theorem shows that a joint posterior profile is implementable by a mediator's public signal if and only if these likelihood ratios satisfy two multiplicative consistency conditions: \emph{internal consistency} (a cycle condition within each common knowledge component) and \emph{external consistency} (a loop condition generated by the mediator pooling states across components).  Our decomposition separates consistency within the players' common-knowledge components from consistency across components linked by the mediator, building on \citet{Rodrigues-Neto2009} and \citet{Hellman2012}.

We then use the aforementioned characterization to study the geometry of the feasible set, showing that it \emph{need not be convex} even in the absence of incentive constraints or restrictions on the number of signals. A characterization of convexity of the feasible set (for every strictly positive prior) in the single-player case is provided, as well. From an applied perspective, non-convexity can arise in familiar persuasion environments, as illustrated by our technology-adoption example with a privately informed firm. Finally, we show how the same graph-based tools inform problems in information design and belief depolarization, and provide an additive analogue relevant for potential games.

Our results bridge abstract information structures and concrete Bayesian updating.  We show that $\f$  can arise from Bayesian posteriors induced by a signal that the mediator releases.  This opens the door to interpreting the mediator as a generator of Blackwell experiments and leads to new insights about the implementability of distributions over states and the coherence of beliefs across players. In doing so, we extend the theory of common priors and beliefs, clarify the conditions under which players’ posteriors can be coherently derived from shared signals, and provide a graph-theoretic approach to understanding the flow of information in multi-agent systems.

\heading{Technology adoption with private assessments: motivating example.} \phantomsection\label{Section - intro. example}

A technology provider discloses information about the quality of its technology to a firm that has received a private assessment.\footnote{For a related persuasion model with a privately informed receiver, see \citet{Guo2019}.} The provider observes the quality but not the assessment, and commits to a public reporting rule based on quality. If the firm adopts, it receives a gross return $q\in\{0,1\}$ and pays a fixed implementation cost $k=\tfrac12$, whereas declining yields a net return of zero. The firm adopts when its expected gross return is at least $k$. We refer to adoption as trade and to declining as no trade. The provider obtains a fixed positive benefit from adoption and thus maximizes its probability. In this example, the firm is a single player, and the provider is the mediator.

There are four states $\Omega=\{\omega_1,\omega_2,\omega_3,\omega_4\}$ with prior $\mu=(\tfrac3{10},\tfrac3{10},\tfrac1{10},\tfrac3{10})$. Quality is good ($q=1$) in $\omega_1$ and $\omega_3$, and bad ($q=0$) in $\omega_2$ and $\omega_4$. The private assessment is favorable in $\omega_1$ and $\omega_2$, and unfavorable in $\omega_3$ and $\omega_4$. Thus, the firm's and provider's information partitions are $\mathcal P_1= \bigl\{\{\omega_1,\omega_2\},\{\omega_3,\omega_4\}\bigr\}$ and $F=\bigl\{\{\omega_1,\omega_3\},\{\omega_2,\omega_4\}\bigr\}$, as illustrated in Figure~\ref{fig:intro-nonconvex-example}.

Before the public report, the firm assigns probability $\tfrac12$ to good quality after a favorable assessment and $\tfrac14$ after an unfavorable assessment. The provider can choose what evidence its report conveys about quality, but it cannot tailor the report to the firm's private assessment.

\begin{figure}[ht]
\centering
\begin{tikzpicture}[scale=0.7]

\draw[thick] (0,0) rectangle (6,6);
\node at (0.35,5.7) {$\Omega$};

\filldraw[black] (1.3,4.5) circle (2pt) node[anchor=west] {$\omega_1$};
\filldraw[black] (1.3,1.5) circle (2pt) node[anchor=west] {$\omega_2$};
\filldraw[black] (4.25,4.5) circle (2pt) node[anchor=west] {$\omega_3$};
\filldraw[black] (4.25,1.5) circle (2pt) node[anchor=west] {$\omega_4$};

\draw[blue, thick] (1.55,3) ellipse (0.85 and 2.2);
\draw[blue, thick] (4.55,3) ellipse (0.85 and 2.2);
\node[blue] at (3.1,3.1) {$\mathcal P_1$};

\draw[red, thick] (3.05,4.5) ellipse (2.25 and 0.75);
\draw[red, thick] (3.05,1.5) ellipse (2.25 and 0.75);
\node[red] at (3,5.55) {$F$};

\end{tikzpicture}
\caption{The player's information partition $\mathcal P_1$ (blue) and the mediator's partition $F$ (red).} \label{fig:intro-nonconvex-example}
\end{figure}

The provider's knowledge of the private assessment has a substantial effect on persuasion. If it observes both quality and the assessment, the maximum probability of trade is $\tfrac45$. For example, an adoption report sent with probability one in $\omega_1,\omega_2,\omega_3$ and with probability $\tfrac13$ in $\omega_4$ occurs with probability $\tfrac45$ and makes the firm's posterior probability of good quality equal to $\tfrac12$ after either assessment. The firm therefore adopts whenever this report is issued.

However, when the provider observes only quality, the maximum probability of trade drops to $\tfrac35$. Strikingly, this probability is already attained without disclosure: the firm adopts after a favorable assessment, which occurs with probability $\tfrac35$. Thus, at this cost, disclosure cannot increase the
probability of trade when the provider observes only quality. The information restriction doubles the minimum probability of no trade from $20\%$ to $40\%$. We establish both optima in Section~\ref{Section - How to optimize}.

To understand why this happens, we need to analyze the constraints on conditional beliefs. Let $x$ and $y$ denote the firm's posterior probabilities of good quality after a favorable and an unfavorable assessment, respectively, conditional on a public report. The corresponding beliefs over states are $(x,1-x,0,0)$ and $(0,0,y,1-y)$, so we can simply represent them by $(x,y)$. When the provider observes only quality, the report must have the same likelihood $a$ in both good states and the same likelihood $b$ in both bad states. whenever $a,b>0$, Bayes' rule yields \(x=\frac{a}{a+b}\) and \(y=\frac{a}{a+3b}\), so we get a non-linear relation \(y=\tfrac{x}{3-2x}\). Note that every point on this curve with $0<x<1$ is feasible: choose $a=x$ and $b=1-x$, and complete the experiment with a complementary signal. Figure~\ref{fig:feasible-beliefs} shows two feasible profiles whose average lies off the curve, establishing non-convexity. The fully informed policy's adoption profile $(\tfrac12,\tfrac12)$ also violates this restriction: when $x=\tfrac12$, feasibility requires $y=\tfrac14$.

\begin{figure}[htbp]
\centering
\begin{tikzpicture}
\begin{axis}[
  width=7.5cm, height=5.8cm,
  xmin=0, xmax=1, ymin=0, ymax=1,
  axis lines=left,
  xtick={0,0.25,0.5,0.75,1},
  ytick={0,0.1,0.3,0.5,1},
  xticklabels={$0$,$\tfrac14$,$\tfrac12$,$\tfrac34$,$1$},
  yticklabels={$0$,$\tfrac1{10}$,$\tfrac3{10}$,$\tfrac12$,$1$},
  xlabel={Favorable assessment, $x$},
  ylabel={Unfavorable assessment, $y$},
  label style={font=\small},
  ticklabel style={font=\small},
  clip=false
]
\addplot[black, semithick, domain=0:1, samples=150]  {x/(3-2*x)};
\addplot[black, dashed, thin]  coordinates {(0.25,0) (0.25,0.1) (0,0.1)};
\addplot[black, dashed, thin] coordinates {(0.5,0) (0.5,0.3) (0,0.3)};
\addplot[black, dashed, thin] coordinates {(0.75,0) (0.75,0.5) (0,0.5)};
\addplot[black, thin]  coordinates {(0.25,0.1) (0.75,0.5)};
\addplot[black, only marks, mark=*, mark size=2pt] coordinates {(0.25,0.1) (0.75,0.5)};
\addplot[black, only marks, mark=x, mark size=3pt, semithick] coordinates {(0.5,0.3)};

\node[anchor=south east, font=\small] at (axis cs:0.235,0.12) {$s_1$};
\node[anchor=south east, font=\small] at (axis cs:0.73,0.52) {$s_2$};
\node[anchor=south east, font=\footnotesize] at (axis cs:0.55,0.31) {Average};
\node[anchor=south east, font=\small] at (axis cs:0.88,0.72) {$y=\tfrac{x}{3-2x}$};
\end{axis}
\end{tikzpicture}
\caption{Feasible profiles and their arithmetic average. In the binary experiment $\tau$, report $s_1$ occurs with probability $\tfrac14$ in both good states and $\tfrac34$ in both bad states; $s_2$ is the complementary report.}
\label{fig:feasible-beliefs}
\end{figure}

\heading{The characterization.} The example raises two questions: which joint posterior profiles can a fixed mediator generate, and when is their feasible set convex? Our analysis answers the first question for any finite state space and any number of players, and characterizes convexity in the single-player case.

The two private information cells in the example are its \emph{common knowledge components} (CKCs). The mediator links them through the pairs $\{\omega_1,\omega_3\}$ and $\{\omega_2,\omega_4\}$, creating an \emph{$F$-loop}, $((\omega_1,\omega_2),(\omega_4,\omega_3))$. Along this loop, the product of signal likelihood ratios must equal one. For the example's prior, this requirement is equivalent to $y=x/(3-2x)$. We refer to this requirement as \emph{external consistency}.\footnote{The loop-product restriction also appears in \citet{Lagziel2025d}, who compare information providers. Here, we fix the mediator and characterize implementable joint posterior profiles.}

In general, we also require \emph{internal consistency}: the product of likelihood ratios must equal one along every $F$-cycle within a CKC. An $F$-cycle follows pairs of states that a player cannot distinguish and closes through a pair that the mediator cannot distinguish. Together with measurability of the target's support, these conditions provide our characterization of implementable joint posteriors.

Using this characterization, we study the geometry of the feasible set. In the single-player setting, it is convex for every strictly positive prior if and only if there is no fully informative $F$-loop (Theorem~\ref{Th:convexity}). The introductory example illustrates how such a loop can produce a non-convex feasible set.  Before optimizing a signaling policy, one must account for these restrictions on the profiles that its signals can generate.  Section~\ref{Section - non convex set of posteriors} returns to the example using the formal consistency conditions and then develops the general convexity result.

\heading{Related literature.}  Several strands of the literature inform our analysis. Since the foundational work of \cite{harsanyi1967games},  much effort has been devoted to formalizing and understanding the informational structures that underlie such environments. A key insight in this literature is that players' information can be represented as partitions over a finite state space, and the relationships among these partitions encode the possible flow and structure of information in the game. \cite{harsanyi1967games} provides the basis for belief-based reasoning in strategic settings.

Aumann’s framework for knowledge and common knowledge (see \citealp{Aumann1976}) formalized the use of partitions to represent agents’ information. Building on this foundation, our paper extends the idea by introducing an external mediator that generates signals constrained by its own partition. A central concept in our analysis is the CKC, which is rooted in Aumann’s original formulation.

The question of whether a joint posterior originates from a \emph{common prior} has attracted significant attention across various settings. This inquiry has been studied extensively in contexts ranging from finite state spaces (see, e.g., \citealp{morris1994trade}), to compact state spaces (e.g., \citealp{feinberg2000characterizing, heifetz2006positive}), and to countable state spaces (see \citealp{lehrer2014belief}). In the present framework, we assume that players initially share a common prior and acquire information through their individual partitions of the state space, subsequently updating their beliefs via Bayesian conditioning. 

The novel aspect of our model is the introduction of an external information source, the mediator.  Given the observed joint posterior profile of the players, we inquire whether such a profile can be rationalized as arising from a known common prior, augmented by additional public information disseminated by the mediator, which itself may be only partially informative to the players.  In other words, we ask: \emph{does the joint posterior stem from the interaction between a common prior and an external, symmetric informational input?}

The conditions we impose, internal and external consistency, are inspired by the cycle-based consistency concepts introduced by \cite{Rodrigues-Neto2009} and \cite{Hellman2012}, though our focus is on implementability via constrained signals rather than belief structures. 

At a conceptual level, our mediator corresponds to a restricted Blackwell experiment \cite{blackwell1951}, with implementability shaped by its coarse knowledge. In this sense, our work complements the Bayesian persuasion literature by asking when a mediator can induce a given belief under informational constraints. 

\cite{Lagziel2026a} compare deterministic information providers, characterizing dominance across games through simultaneous matching of all players’ posterior partitions. The present paper fixes the mediator’s information and characterizes the joint posterior profiles attainable through stochastic public signaling.

Motivated by the question of aggregating experts' opinions and building on the results of \cite{Kellerer1961}, \cite{Strassen1965}, and \cite{Gutmann1991}, the work of \cite{Dawid1995} provides a characterization of joint posterior beliefs for the case of two agents. This was extended by \cite{Arieli2021}, who characterize the set of feasible distributions of joint posterior beliefs that can arise among multiple agents in a binary state space, given a Blackwell experiment that provides different (potentially asymmetric) information to each agent. Their characterization is closely related to the no-trade literature in that it identifies constraints on belief distributions that are consistent with a common prior and Bayesian updating, even when agents receive heterogeneous private signals.\footnote{See also the follow-up paper and review of the no-trade history in \cite{Morris2020}, as well as the study of \cite{Burdzy2020}, which follows \cite{Dawid1995}, to derive probabilistic bounds on the polarization of posteriors in a two-agent setting.} More recently, \cite{Herings2024} studied a similar question, but derived a different characterization while accounting for an arbitrary finite state space. 

Independently of \cite{Arieli2021}, \cite{Ziegler2020} studies a setting in which a mediator provides private signals to two receivers without committing to a common information structure. The mediator therefore chooses information structures robustly, maximizing expected payoff against the worst-case interpretation consistent with Bayesian rationality.  In doing so, Ziegler derives necessary feasibility constraints on the joint distribution of receivers' posteriors, conditions that coincide with those of \citet{Arieli2021} as both necessary and sufficient in the case of two agents.

Our analysis distinguishes itself from existing literature in four fundamental ways, which collectively define a novel research agenda.  First, we adopt a setting in which players receive private signals from a fixed information structure regarding an unknown state that is not necessarily binary, allowing for a richer initial belief space. Second, the mediator is only partially informed about the true state, and their own information structure is fixed, which shifts the focus from optimal information design to the mechanics of communication under given constraints.  Our characterization is thus framed in terms of the information structures of both the players and the mediator. Third, the mediator utilizes a public communication channel, a deliberate simplification that allows us to focus on the impact of generating common knowledge, rather than the complexities of personalized private signals.  Finally, and most critically, our primary focus is on characterizing the specific joint posteriors that can be induced by the mediator, rather than the distribution over these posteriors.   We then build upon this to define and characterize the entire set of implementable joint posteriors, investigating the conditions under which any element of this set can be implemented through a stochastic public signal, a major departure from the standard distributional analysis in this field.

These differences become most apparent in our final result, which shows that convexity (of the set of joint posterior beliefs) can fail purely due to information constraints.  While instances of non-convexity are well known, they typically stem from different sources.  For example, \cite{Herings2024} obtain non-convexity of feasible posterior distributions under limited message alphabets, whereas our framework imposes no such restriction.  \cite{Candogan2023} derive a finite-dimensional formulation of multi-agent information design that need not be convex, whereas our non-convexity concerns the feasible set of conditional joint posterior profiles and is entirely payoff-free. \cite{Chen2025} demonstrate non-convex feasible regions under support-size caps (K-signaling), whereas in our setting, constraints emerge from likelihood-ratio products on F-loops, without any bound on the support.  \cite{Kosenko2020} visualize non-convex feasible-posterior shapes under garbling, while we provide necessary and sufficient graph-based conditions, identifying precisely when such non-convexity arises from a partially informed mediator and privately informed players.  Finally, \cite{Ball2021} establish non-convexity when the set of experiments is restricted, whereas in our framework, the restriction is due to the measurability of kernels given the mediator’s information. 

Our characterization also clarifies when convexity can and cannot be expected.  In Section~\ref{Section - non convex set of posteriors} we provide a sharp convexity characterization for the single-player case, and we show by example that non-convexity can persist in richer environments, including settings with multiple privately informed players, even when the mediator is fully informed. Together, these findings build the case that non-convexity can arise purely from informational constraints, independent of payoffs, support, or signal limitations.

Finally, our results contribute to the broader literature on higher-order beliefs (see the review by \citealp{Geanakoplos1994}), by characterizing which belief patterns, encoded in posterior likelihoods, can emerge from shared public signals constrained by a third party’s limited information.

\heading{The structure of the paper.} \label{Section - Structure} The paper is organized as follows.  Section~\ref{Section - Preliminaries} presents the model and defines joint beliefs, posterior likelihoods, the graph of information, and the notions of cycles and loops.  Section~\ref{Section - Consistency condition and main result} introduces internal and external consistency and proves the key graph-theoretic result (Proposition~\ref{Proposition: basic charecterization}) underlying our main characterization.  Section~\ref{Section: Main characterization} states the main implementability result (Theorem~\ref{th: main Bayesian}). 
Section~\ref{Section - non convex set of posteriors} studies the convexity of the feasible set. 
Section~\ref{Section - Extensions and applications} discusses extensions and applications, including multiple signals, persuasion, polarization, and potential games.

\section{Preliminaries} \label{Section - Preliminaries}

Let $N = \{1, 2, \dots, n\}$ denote the set of players, and let $\Omega$ be a non-empty, finite state space with a strictly positive common prior $\mu$. Each player $i \in N$ has a finite partition $\mathcal{P}_i$ of $\Omega$, representing player $i$'s information.   For any state $\omega \in \Omega$, we denote by $\mathcal{P}_i(\omega)$ the element of the partition $\mathcal{P}_i$ that contains $\omega$. A Common Knowledge Component (CKC), typically denoted by $C \subseteq \Omega$, is a minimal non-empty subset of states that is measurable with respect to the $\sigma$-algebra of every player (see \citealp{Aumann1976}). The notation $C(\o)$ refers to the CKC that contains the state $\o$.

Let $F$ be a partition of $\Omega$ belonging to an agent outside the set $N$, which we refer to as the \emph{mediator}.  As before, $F(\o)$ denotes the information set of $F$ that contains $\o.$  The mediator may provide additional information to the players beyond their private information.  For this purpose, the mediator uses a public signaling function $\tau: \Omega \to \Delta (S)$, where $S$ is a finite set of signals and  $\Delta (S)$ is the set (simplex) of distributions over $S$. Let $\tau(s \mid \omega)$ denote the probability that the public signal $s \in S$ is observed given state $\omega$. Note that the signaling function is measurable with respect to (henceforth, w.r.t.) $F$.\footnote{Measurability w.r.t.\ $F$ means that if $\o$ and $\o'$ are indistinguishable from the mediator's perspective, namely, in case $\o' \in F(\o)$, then  $\tau(s\mid \o')= \tau(s\mid \o)$.}
Formally, $\tau$ is an $F$-measurable stochastic kernel:
\begin{itemize}
  \item[(i)] For each state $\omega \in \Omega$, $\tau(\cdot \mid \omega)$ is a probability distribution over a finite set of signals $S$.
  \item[(ii)] For all $\omega' \in F(\omega)$ and $s \in S$, we have $\tau(s \mid \omega') = \tau(s \mid \omega)$.
\end{itemize}
The signaling function $\tau$ is also known as a Blackwell experiment (see \citealp{blackwell1951}), so the mediator can be viewed as a generator of Blackwell experiments.

\subsection{Bayesian updating and joint posterior beliefs} \label{sec: posteriors}

When the realized state is ~$\omega$, player $i$ is informed of $\mathcal{P}_i(\omega)$.  Assuming there exists $\o'\in \mathcal{P}_i(\o)$ such that $\tau(s|\o')>0$, upon observing $s$, player~$i$ updates his belief using Bayes' rule.  The updated belief is given by the following distribution.
\begin{equation} \label{eq: i'th posterior}
    \mu_{\tau,i}(\omega' \mid \mathcal{P}_i(\omega), s) =
    \frac{\mathbf{1}_{\{\omega'\in\mathcal{P}_i(\omega)\}}\mu(\omega')\tau(s\mid \omega')}
    {\sum_{\omega''\in \mathcal{P}_i(\omega)}\mu(\omega'')\tau(s\mid \omega'')}.
\end{equation}
In particular, $\mu_{\tau,i}(\cdot\mid \mathcal{P}_i(\omega), s)$ is a probability distribution whose support is a subset of $\mathcal{P}_i(\omega)$.   For notational convenience, we set $\mu_{\tau,i}(\omega' \mid \mathcal{P}_i(\omega), s) = 0$ for every $\o'$, whenever $\tau(s \mid \mathcal{P}_i(\omega)) = 0$.

When the realized state may vary, we define the \emph{joint posterior associated with $\tau$ and $s$} as the set of posterior profiles across all states and players that can be generated by $s$:
\begin{equation}\label{eq: the joint posterior of mu , s}
  \boldsymbol{\mu}_{\tau,s}
=\Bigl\{\,\bigl(\mu_{\tau,i}(\cdot \mid \mathcal{P}_i(\omega),s)\bigr)_{i\in N} \;:\; \omega\in\Omega \text{ with } \tau(s\mid\omega)>0 \Bigr\}.
\end{equation}
This joint posterior dictates each player’s beliefs about the true state, as well as their beliefs about others’ beliefs and higher-order beliefs, making it a central object for analyzing equilibrium behavior.

\subsection{Joint beliefs} \label{Section - joint beliefs}

To allow for even greater generality, the following definition of a \emph{joint belief} requires neither a signal nor a mediator.  It will be used throughout the paper to define the joint profile of beliefs that a mediator can potentially generate. 

\begin{definition}\label{def: Joint posterior}
A \emph{joint belief} is a map $\emph{JB}:\Omega\times N \to \Delta(\Omega)\cup\{\1_{\emptyset}\}$,\footnote{The term $\emph{\1}_{\emptyset}$ denotes the zero vector.} not identically equal to $\1_{\emptyset}$, satisfying:
\begin{enumerate}[\emph{(}i\emph{)}]
    \item \emph{Common realized state}: if $\emph{JB}(\omega,i)(\omega)>0$ for some $i$, then $\emph{JB}(\omega,j)(\omega)>0$ for all $j\in N$.
    \item \emph{Information-set support}: if $\emph{JB}(\omega,i)\neq \1_{\emptyset}$, then the support of $\emph{JB}(\omega,i)$ is a subset of $\mathcal{P}_i(\omega)$, and $\emph{JB}(\omega,i)(\omega)>0$. Moreover, for every $\omega'\in\Omega$, $\emph{JB}(\omega,i)(\omega')>0$ implies $\emph{JB}(\omega',i)\neq\1_{\emptyset}$.
    \item \emph{Information-set invariance}: for every $\omega'\in \mathcal{P}_i(\omega)$, if $\emph{JB}(\omega,i)\neq \1_{\emptyset}$ and $\emph{JB}(\omega',i)\neq \1_{\emptyset}$, then $\emph{JB}(\omega',i)=\emph{JB}(\omega,i)$.
    \item \emph{Likelihood-ratio agreement}: for every $i,j,\o$ and $\o' \in \mathcal{P}_i(\omega) \cap \mathcal{P}_j(\omega)$,
    \[
    \frac{\emph{JB}(\omega,i)(\omega)}{\emph{JB}(\omega',i)(\omega')} = \frac{\emph{JB}(\omega,j)(\omega)}{\emph{JB}(\omega',j)(\omega')},
    \]
    assuming the denominators are nonzero.
\end{enumerate}
\end{definition}

When well-defined (i.e., when ${\rm{JB}}(\omega,i)\neq \emph{\1}_{\emptyset}$), ${\rm{JB}}(\omega,i)$ stands for the belief distribution that player $i$ assigns to $\Omega$ when the set $\mathcal{P}_i(\omega)$ is realized. The corresponding probability of state $\omega'$ is denoted by ${\rm{JB}}(\omega,i)(\omega')$.

Several remarks are in order. First, the conditions in Definition~\ref{def: Joint posterior} are necessary for a joint belief to be induced as a joint posterior belief by a public signal. Indeed, they follow from the basic properties of Bayesian inference given the existence of a common prior and a mediator's public signal.\footnote{A broader discussion of this point is provided in connection with Eq.~\eqref{eq: phi is defined here} below.} For example, a common realized state (condition (i) above) must be satisfied whenever players consider the same probability space, and in cases without a common prior, likelihood-ratio agreement (condition (iv) above) need not be satisfied. Second, we include the indicator $\emph{\1}_{\emptyset}$ to capture cases where the mediator's signal has zero likelihood at $\omega$; in such cases, we set ${\rm{JB}}(\omega,i)=\emph{\1}_{\emptyset}$ for all such states. Note that $\tau(s\mid \omega')=0$ for some $\omega'\in \mathcal{P}_i(\omega)$ does not by itself make the posterior in Eq.~(1) ill-defined; it only implies that $\mu_{\tau,i}(\omega'\mid \mathcal{P}_i(\omega),s)=0$. Third, the definition hinges on the players' fixed partitions, primarily through the information-set support and invariance requirements (conditions (ii) and (iii) above), but it does not relate to the mediator's partition. Moreover, these two conditions apply to each player \emph{at the individual level}, rather than across players.

Given a JB and a common prior $\mu$, a natural question is whether there exists a signaling function $\tau$ and a signal $s$ that jointly induce this JB as a posterior. We approach this question by examining posterior likelihood ratios. This builds on a key observation that, although posterior updating is typically player-dependent (as the denominator reflects player~$i$'s information partition), the following likelihood ratios are not:
\begin{equation}\label{eq: phi is defined here}
  \frac{ \mu_{\tau,i}(\omega \mid \mathcal{P}_i(\omega), s)}{\mu_{\tau,i}(\omega' \mid \mathcal{P}_i(\omega'), s)} = \frac{\mu(\omega)\tau(s\mid \omega)}{\mu(\omega')\tau(s\mid \omega')}, \ \ \forall \o, \o'\in \mathcal{P}_i(\omega).
\end{equation}
In particular, whenever $\omega$ and $\omega'$ lie in the same information set (i.e., $\omega' \in \mathcal{P}_i(\omega)$, or equivalently $\mathcal{P}_i(\omega) = \mathcal{P}_i(\omega')$) and both conditional probabilities are positive, this ratio is well defined and independent of the player.  This also clarifies the two constraints imposed in Definition \ref{def: Joint posterior}. We explore this idea further in the following section.

To formally define the problem, fix a JB and a common prior $\mu$, and let $\Omega_+$ denote the set of states such that $\text{JB}(\omega,i)(\omega) > 0$ for some player $i$, and hence for all players.  In particular, we restrict attention to the states that are assigned positive probability. Normalizing $\mu$ to this set, we obtain the distribution $\mu(\cdot \mid \O_+)$.\footnote{We discuss the case where some states are assigned zero probability in Section \ref{Section: Main characterization}.}

For a pair $\o,\o'\in \O_+$ such that $\o'\in \mathcal P_i(\o)$ for some $i$, consider the ratio $\tfrac{{\rm{JB}}(\omega,i)(\o)}{{\rm{JB}}(\omega',i)(\o')}$. We ask whether there exists a signaling function $\tau$ of the mediator and a signal $s$ that induce the same likelihood ratios.  Specifically, does there exist a signal $s$ and a function $\tau$ such that 
\begin{equation}\label{eq:likelihood}
    \frac{{\rm{JB}}(\omega,i)(\o)}{{\rm{JB}}(\omega',i)(\o')} = 
     \frac{\mu(\omega\mid \O_+)\tau( s \mid \o )}{\mu(\omega'\mid \O_+)\tau( s \mid \o')}=                                            
    \frac{\mu(\omega)\tau( s \mid \o )}{\mu(\omega')\tau( s \mid \o')},
\end{equation}
for every player~$i$ and every $\omega' \in \mathcal{P}_i(\omega)$? Note that the right-hand side of Eq.~\eqref{eq:likelihood} does not depend on player~$i$, but only on whether $\omega$ and $\omega'$ lie in the same information set for some player. To further discuss this question and see why likelihood ratios are sufficient statistics for generating joint posteriors, consider the following example.

\begin{example} \label{ex:  5 states 2 players}
\rm
There are five states and two players such that $\mathcal{P}_1=\{\{\o_1,\o_2\},\{\o_3\},\{\o_4,\o_5\}\}$, $\mathcal{P}_2=\{\{\o_1,\o_2,\o_3\},\{\o_4\},\{\o_5\}\}$, and $F=\{\{\o_1,\o_4\},\{\o_2,\o_3,\o_5\}\}$. Figure \ref{fig:example of players information} illustrates the knowledge structures of the players as well as that of the mediator. 

\begin{figure}[ht]
\centering
\begin{minipage}{0.45\textwidth}
\centering
\begin{tikzpicture}[scale=0.75]

\draw[thick] (0,0) rectangle (6.6,6);

\node at (0.3,5.7) {$\Omega$};

\node[blue] at (3.5,4) {$\mathcal{P}_1$};

\node[red] at (2.9,2) {$\mathcal{P}_2$};

\draw[red, thick] (1.5,3) ellipse (1 and 2.7);
\draw[blue, thick] (1.5,3.9) ellipse (0.8 and 1.5);
\draw[blue, thick] (1.5,1.5) ellipse (0.7 and 0.7);

\draw[blue, thick] (4.9,3) ellipse (1 and 2.7);
\draw[red, thick] (4.9,4.5) ellipse (0.7 and 0.7);
\draw[red, thick] (4.9,1.5) ellipse (0.7 and 0.7);

\filldraw[black] (1.35,4.5) circle (2pt) node[anchor=west] {$\o_1$};
\filldraw[black] (1.35,3.2) circle (2pt) node[anchor=west] {$\o_2$};
\filldraw[black] (1.35,1.5) circle (2pt) node[anchor=west] {$\o_3$};
\filldraw[black] (4.8,4.5) circle (2pt) node[anchor=west] {$\o_4$};
\filldraw[black] (4.8,1.5) circle (2pt) node[anchor=west] {$\o_5$};

\end{tikzpicture}
\caption*{A}
\vspace{-0.2cm}
\caption*{The players' information}
\end{minipage}%
\hfill
\begin{minipage}{0.45\textwidth}
\centering
\begin{tikzpicture}[scale=0.75]

\draw[thick] (0,0) rectangle (6.6,6);

\node at (0.3,5.7) {$\Omega$};

\node[orange] at (3.5,5.6) {$F$};

\draw[orange, thick, rotate around={90:(3,3)}] (4.5,2.5) ellipse (0.8 and 2.7);

\coordinate (A) at (1.3,0.8);
\coordinate (B) at  (1.3,3.7);
\coordinate (C) at (6.2,1.2) ;

\draw[orange, thick, smooth cycle] plot coordinates {(A) (B) (C) };

\filldraw[black] (1.4,4.5) circle (2pt) node[anchor=west] {$\o_1$};
\filldraw[black] (1.4,3) circle (2pt) node[anchor=west] {$\o_2$};
\filldraw[black] (1.4,1.5) circle (2pt) node[anchor=west] {$\o_3$};
\filldraw[black] (4.9,4.5) circle (2pt) node[anchor=west] {$\o_4$};
\filldraw[black] (4.9,1.5) circle (2pt) node[anchor=west] {$\o_5$};

\end{tikzpicture}
\caption*{(B)}
\vspace{-0.2cm}
\caption*{The mediator's information}
\end{minipage}
\caption{Panel A illustrates the information structure of player $1$ (blue) and player $2$ (red). Panel B portrays the information structure of the mediator (orange).}
\label{fig:example of players information}
\end{figure}

\end{example}

  Starting with a basic set-up and in the absence of any mediator, one may ask whether there exists a common prior that induces the joint posterior given in Table~\ref{tab: 1}. 

\begin{table}[H]
    \centering
    \caption{Joint beliefs without a mediator}
    \label{tab: 1}
    \footnotesize

    \begin{tabular}{@{}ccc@{}}
        \toprule
        State & Player 1 & Player 2 \\
        \midrule
        $\o_1$
        & $(\nicefrac{1}{2},\nicefrac{1}{2},0,0,0)$
        & $(\nicefrac{1}{3},\nicefrac{1}{3},\nicefrac{1}{3},0,0)$ \\
        $\o_2$
        & $(\nicefrac{1}{2},\nicefrac{1}{2},0,0,0)$
        & $(\nicefrac{1}{3},\nicefrac{1}{3},\nicefrac{1}{3},0,0)$ \\
        $\o_3$
        & $(0,0,1,0,0)$
        & $(\nicefrac{1}{3},\nicefrac{1}{3},\nicefrac{1}{3},0,0)$ \\
        $\o_4$
        & $(0,0,0,\nicefrac{1}{2},\nicefrac{1}{2})$
        & $(0,0,0,1,0)$ \\
        $\o_5$
        & $(0,0,0,\nicefrac{1}{2},\nicefrac{1}{2})$
        & $(0,0,0,0,1)$ \\
        \bottomrule
    \end{tabular}

    \par\smallskip
    \begin{minipage}{\textwidth}
        \footnotesize
        Note: Each row reports the players' posterior beliefs
        conditional on the realized state.
    \end{minipage}
\end{table}

Indeed, such a prior exists: the uniform distribution over $\Omega$ induces exactly these posterior beliefs.  This example also illustrates a restriction across players: the states $\omega_1$ and $\omega_2$ lie in the same information set for both players.

As a consequence, the ratio between the probabilities assigned to these states (namely, $\tfrac{1/2}{1/2}$ for player~1 and $\tfrac{1/3}{1/3}$ for player~2) is identical across players.  This property persists even when a mediator provides additional information, as illustrated in the next table, and it constitutes a cornerstone of our characterization.

Suppose now that the common prior is uniform.  Does there exist a public signal that the mediator could reveal so as to induce the JB given in Table~\ref{tab: 2}?

\begin{table}[H]
    \centering
    \caption{Joint beliefs conditional on the mediator's signal}
    \label{tab: 2}
    \footnotesize

    \begin{tabular}{@{}ccc@{}}
        \toprule
        State & Player 1 & Player 2 \\
        \midrule
        $\o_1$
        & $(\nicefrac{1}{3},\nicefrac{2}{3},0,0,0)$
        & $(\nicefrac{1}{5},\nicefrac{2}{5},\nicefrac{2}{5},0,0)$ \\
        $\o_2$
        & $(\nicefrac{1}{3},\nicefrac{2}{3},0,0,0)$
        & $(\nicefrac{1}{5},\nicefrac{2}{5},\nicefrac{2}{5},0,0)$ \\
        $\o_3$
        & $(0,0,1,0,0)$
        & $(\nicefrac{1}{5},\nicefrac{2}{5},\nicefrac{2}{5},0,0)$ \\
        $\o_4$
        & $(0,0,0,\nicefrac{1}{3},\nicefrac{2}{3})$
        & $(0,0,0,1,0)$ \\
        $\o_5$
        & $(0,0,0,\nicefrac{1}{3},\nicefrac{2}{3})$
        & $(0,0,0,0,1)$ \\
        \bottomrule
    \end{tabular}

    \par\smallskip
    \begin{minipage}{\textwidth}
        \footnotesize
        Note: The common prior is uniform. Each row reports
        the players' posterior beliefs conditional on the
        realized state and the mediator's signal.
    \end{minipage}
\end{table}

The answer is yes. Consider a signaling function $\tau$ and a signal $s$ such that $\tau(s\mid \omega_i)=\nicefrac{1}{5}$ for $i\in\{1,4\}$ and $\tau(s\mid \omega_i)=\nicefrac{2}{5}$ for $i\in\{2,3,5\}$. Since $\tau(s\mid \omega_1)=\tau(s\mid \omega_4)$ and $\tau(s\mid \omega_2)=\tau(s\mid \omega_3)=\tau(s\mid \omega_5)$, the function $\tau$ is $F$-measurable.

Now modify Table~\ref{tab: 2} by replacing player~1's posteriors at $\omega_4$ and $\omega_5$ with $\left(0,0,0,\nicefrac{1}{5},\nicefrac{4}{5}\right)$. The resulting table still defines a joint belief (Definition~\ref{def: Joint posterior}), but it is \emph{not} implementable by any $F$-measurable public signal.

The reason is that if, under $\omega_1$, the posterior of player~1 is $\left(\frac{1}{3}, \frac{2}{3}, 0, 0, 0\right)$, then due to the $F$-measurability restriction, the ratio
$$
\frac{\tau(s\mid \omega_1)}{\tau(s\mid \omega_2)} \quad \text{must match} \quad \frac{\tau(s\mid \omega_4)}{\tau(s\mid \omega_5)}.
$$
However, in the former case, the ratio is $\frac{1}{2}$, while in the latter it is $\frac{1}{4}$.

\heading{A hierarchy of beliefs.} A joint belief specifies each player's posterior at every state, and therefore encodes not only first-order beliefs but also higher-order beliefs, which are central for equilibrium analysis. For example, in Table~\ref{tab: 1}, when $\omega_4$ is realized player~1 assigns probability $\nicefrac{1}{2}$ to each of $\omega_4$ and $\omega_5$. Hence player~1 assigns probability $\nicefrac{1}{2}$ to the event that player~2 is certain the state is $\omega_4$ and probability $\nicefrac{1}{2}$ to the event that player~2 is certain the state is $\omega_5$. This illustrates how the posteriors' state-by-state specification also encodes the \emph{hierarchy of beliefs}.

This example also highlights why likelihood ratios are the right object for our analysis: any $F$-measurable public signal must preserve relative likelihoods of states within each information set, and these ratios will become the algebraic constraints in our implementability characterization.

\subsection{The graph of information and PL functions}  \label{Section - The Graph and the Posterior Likelihood function}

In this section, we present two key elements for our analysis: the graph of information and posterior likelihood functions. Starting with the former, we represent the agents' private information by a directed graph, following \cite{Rodrigues-Neto2009}.  The graph provides a compact way to record which pairs of states are \emph{directly comparable} for at least one player (i.e., states that a player may confound within an information set).  This will later allow us to express Bayesian feasibility restrictions as simple multiplicative constraints along particular chains in the graph.

Formally, let $G=(V,E)$ be a directed graph with vertex set $V=\Omega$.  For two states $\omega,\omega'\in\Omega$, we include a directed edge $(\omega,\omega')\in E$ (denoted $\omega\to\omega'$) if there exists a player $i\in N$ such that $\omega'\in \mathcal{P}_i(\omega)$.  Since each $\mathcal{P}_i$ is a partition, whenever $\omega\to\omega'$ we also have $\omega'\to\omega$.  We keep both directions because likelihood ratios will be defined on directed edges. See below the illustrative graph for Example \ref{ex:  5 states 2 players}.

Let $\twoheadrightarrow$ denote the transitive closure of $\to$: we write $\omega \twoheadrightarrow \omega'$ if there exists a finite sequence of states $\omega = \omega_1,\omega_2,  \ldots,\omega_{m} = \omega'$ such that $\omega_k\to\omega_{k+1}$ for all $k<m$. Because edges come in both directions, $\twoheadrightarrow$ is an equivalence relation; its equivalence classes coincide with the common knowledge components (CKCs). In particular, $G$ decomposes into connected components, each corresponding to a CKC.

The second central object in our characterization is a function that records \emph{local} likelihood ratios (comparisons across edges).
\begin{definition}[Posterior likelihood (PL) function]
    A positive function $\f$ defined over $E$ is a \emph{posterior likelihood} function if it satisfies the condition\footnote{We abuse notation and use $\f(\omega, \omega')$ instead of $\f((\omega, \omega'))$.} 
\begin{equation} \label{eq:phi_reversed}
 \f(\omega, \omega')=\frac{1}{\f(\omega', \omega)}.
\end{equation}
\end{definition}

Intuitively, $\f(\omega,\omega')$ is the likelihood ratio of $\omega$ relative to $\omega'$ as perceived by any player who cannot distinguish between them.\footnote{A PL function is a gain function, as defined in \citet{Zaslavsky1989}, with values in the multiplicative group of positive real numbers; our terminology emphasizes its Bayesian interpretation.} Later, a joint belief (together with a prior) will induce such an $\f$ on $E$ via Bayes' rule. We typically ignore self edges (as $\o\in \mathcal{P}_i(\o)$ trivially), but if needed, set $\f(\o,\o)=1$.

\noindent {\bf{Example \ref{ex:  5 states 2 players}, continued.}} 
The graph of information corresponding to the model described in Example \ref{ex:  5 states 2 players}, omitting arrows from a state to itself, is given in Figure \ref{fig:the graph of information}. 
Note that the likelihood ratio of the probabilities of $\o_1$ and  $\o_2$ is $\nicefrac{1}{2}$ (recall Table \ref{tab: 2} above). 
This is true for both players since $\o_1$ and  $\o_2$  belong to the same information sets of the two.
We therefore obtain that $\f(\omega_1, \omega_2) = \nicefrac{1}{2}$,  and the other values of the PL function are: 
\begin{equation}\label{eq: phi in table 2}
\f(\omega_2, \omega_3) = 1, \quad
\f(\omega_3, \omega_1) = 2, \quad \text{and} \quad \f(\omega_4, \omega_5) = \nicefrac{1}{2},
\end{equation}
while preserving the relation in Eq.~\eqref{eq:phi_reversed}.

\begin{figure}[th!]
\centering
\begin{minipage}{0.45\textwidth}
\centering
\begin{tikzpicture}[scale=1]

\filldraw[black] (1.5,4.5) circle (2pt) node[anchor=west] {$\o_1$};
\filldraw[black] (1.5,1.5) circle (2pt) node[anchor=east] {$\o_2$};
\filldraw[black] (3.5,1.5) circle (2pt) node[anchor=west] {$\o_3$};
\filldraw[black] (5.3,4.5) circle (2pt) node[anchor=west] {$\o_4$};
\filldraw[black] (5.3,1.5) circle (2pt) node[anchor=west] {$\o_5$};

\draw (1.5,4.5)-- (1.5,1.5) --(3.5,1.5) -- (1.5,4.5);
\draw (5.3,4.5)-- (5.3,1.5);
\end{tikzpicture}

\vspace{-0.2cm}
\end{minipage}%

\caption{The graph of information. In this example, there are two connected components, each corresponding to a CKC.}
\label{fig:the graph of information}
\end{figure}

\subsection{Cycles and information loops}\label{Section: cycles-loops}

The mediator's partition $F$ may be coarser than the players' common-knowledge partition:  the mediator may pool together states that all players (collectively) do distinguish.  This mismatch is captured by two graph-theoretic objects: $F$-cycles and $F$-loops. We introduce them here so that the subsequent consistency section can focus purely on restrictions on $\f$ rather than on the combinatorics of the environment.

An $F$-cycle is a chain of player-edges that ``closes'' because the terminal state lies in the same mediator cell as the initial state.
\begin{definition}[$F$-cycle]\label{def:fcycle}
An \emph{$F$-cycle} is a finite sequence of states $\omega_1,\omega_2,\ldots,\omega_{m+1}$ such that
\[
(\omega_i,\omega_{i+1})\in E \quad \text{for }i=1,\ldots,m,
\qquad\text{and}\qquad
\omega_{m+1}\in F(\omega_1).
\]
\end{definition}
Since $\omega_1 \twoheadrightarrow \omega_{m+1}$, every $F$-cycle is contained in a single CKC; the mediator enters only through the closing condition $\omega_{m+1}\in F(\omega_1)$.  When $F$ is discrete (singleton cells), an $F$-cycle reduces to an ordinary directed cycle $\omega_{m+1}=\omega_1$.

Next, we define information loops that capture the mediator's ability to connect distinct CKCs through its own coarser information. The concept of a loop plays a pivotal role in \cite{Lagziel2025d} to provide conditions such that one mediator dominates another, in an extension of Blackwell's work on the comparison of experiments (see \citealp{blackwell1951}).

\begin{definition}[$F$-loop]\label{def:floop}
An \emph{$F$-loop} is a finite sequence of pairs $\bigl((\omega_i,\bar\omega_i)\bigr)_{i=1}^m$ such that, for each $i=1,\ldots,m$:
\begin{enumerate}[\emph{(}i\emph{)}]
\item $\omega_i \twoheadrightarrow \bar\omega_i$ (the pair lies in the same CKC);
\item $\bar\omega_i \not\twoheadrightarrow \omega_{i+1}$ (successive pairs lie in different CKCs);
\item $\bar\omega_i \in F(\omega_{i+1})$, with the convention $\omega_{m+1}:=\omega_1$.
\end{enumerate}
\end{definition}

Condition (i) means that each pair $(\omega_i,\bar\omega_i)$ lies within a single CKC. Condition (ii) forces the loop to move across CKCs.  Condition (iii) explains \emph{how} it moves: the mediator cannot distinguish $\bar\omega_i$ from the next state $\omega_{i+1}$ because they lie in the same $F$-cell.  In the introductory example, the single player's information cells $\{\omega_1,\omega_2\}$ and $\{\omega_3,\omega_4\}$ are the two CKCs.  The mediator pools $\omega_1$ with $\omega_3$ and $\omega_2$ with $\omega_4$, although the player distinguishes the two CKCs.   The pairs $((\omega_1,\omega_2),(\omega_4,\omega_3))$ therefore form an $F$-loop.

The graph, cycles, and loops introduced above are purely structural objects determined by the players' information partitions and the mediator's partition $F$.  In Section~\ref{Section - consistency conditions}, we translate these objects into algebraic restrictions on the posterior likelihood function $\f$ by evaluating products of likelihood ratios along them. Along an $F$-cycle we multiply $\f$ directly on successive edges, whereas along an  $F$-loop we must compare states within a CKC that may not be adjacent; we do so by extending $\f$ along paths inside each CKC (internal consistency will guarantee that this extension is well defined). These consistency conditions will then be used in our characterization of Bayesian implementability.

\section{Consistency conditions and main result}\label{Section - Consistency condition and main result}

In this section we introduce two restrictions, referred to as \emph{internal} and \emph{external} consistency, that characterize when a posterior-likelihood function $\f$ can be generated by an $F$-measurable signaling function. Section~\ref{Section - consistency conditions} develops the graph-theoretic characterization, and Section~\ref{Section: Main characterization} applies it to Bayesian updating to obtain our main implementability theorem.

\subsection{Internal and external consistency} \label{Section - consistency conditions}

We start with a purely structural question that depends only on the players' partitions $(\mathcal{P}_i)_i$ and the mediator's partition $F$. Fix a PL function $\f$ on the edge set $E$. Does there exist an $F$-measurable, strictly positive function $f:\O\to \mathbb{R}_{++}$ such that
\[
\f(\o,\o')=\frac{f(\o)}{f(\o')} \qquad\text{for every }(\o,\o')\in E?
\]

To address this question, recall the notions of an $F$-cycle and an $F$-loop from Section~\ref{Section: cycles-loops}. We use these objects to formulate two consistency conditions, which Proposition~\ref{Proposition: basic charecterization} shows are necessary and sufficient. This result will later allow us to test whether a joint belief JB is implementable: we map JB into its induced PL function and verify the consistency conditions (see Eq.~\eqref{eq:likelihood} and Remark~\ref{rmk: phi induces probability}).

The first property, \emph{internal consistency}, parallels Definition~2 in \citet{Rodrigues-Neto2009}, where cycles are defined solely in terms of the players' information, independently of the mediator.\footnote{In specific cases, \citet{Hellwig2013} simplifies the consistency test of \citet{Rodrigues-Neto2009} for verifying whether a given set of players' posteriors is compatible with a common prior.}

\begin{definition} \label{Definition: a cycle}
Let $(\o_i, \o_{i+1})_{i=1}^m$ be an $F$-cycle.  Then, \emph{Internal consistency} $\rm\bf[INC]$ holds if
\begin{equation}\label{eq: type ratio}
    \prod_{i=1}^m \f(\o_i, \o_{i+1})=1.   
\end{equation}
\end{definition} 

\citet{Hellman2012} refer to the left-hand side of Eq.~\eqref{eq: type ratio} (when $F$ is trivial) as the \emph{type ratio of a chain}\footnote{More precisely, our induced PL function uses the inverse of their posterior ratio, corrected by prior odds; see Eq.~\eqref{eq: defining phi}.} Note that for every $(\o_1,\o_2)\in E$, the sequence $(\o_1,\o_2,\o_1)$ is an $F$-cycle. Thus, by Eq.~\eqref{eq:phi_reversed}, a positive function $\f$ defined on $E$ is a PL function if and only if Eq.~\eqref{eq: type ratio} holds for every such $F$-cycle. In particular, \textbf{[INC]} implies Eq.~\eqref{eq:phi_reversed}, and hence $\f$ is indeed a PL function.

The following Lemma \ref{lemma:extension} uses $\rm\bf[INC]$ to extend any $\f$ in a consistent manner in every CKC.
(The proof is relegated to the appendix; see Section \ref{Appendix: proof of extension Lemma}.)

\begin{lemma} \label{lemma:extension} 
Assume $\rm\bf[INC]$. 
Then, $\f$ can be extended to any pair  $(\o_1,\o_{m+1})$ where $\o_1 \twoheadrightarrow \o_{m+1}$, and it holds that for any sequence of edges $\bigl((\o_i, \o_{i+1})\bigr)_{i=1}^m$ in $E$,
\begin{equation} \label{eq: External con}
\prod_{i=1}^m \f(\o_i, \o_{i+1}) =\f(\o_1, \o_{m+1}).
\end{equation}
\end{lemma}

In Example \ref{ex:  5 states 2 players}, the sequence $\bigl((\omega_1, \omega_2), (\omega_2, \omega_3), (\omega_3, \omega_1)\bigr)$ forms an $F$-cycle, and indeed, in Table~\ref{tab: 2}, the product of the corresponding values that $\f$ assigns to these edges does equal $1$ (see Eq.~\eqref{eq: phi in table 2} above).

\begin{remark}  \label{rmk: phi induces probability}
\emph{Lemma~\ref{lemma:extension}} implies that if $\f$ satisfies  \emph{\textbf{[INC]}}, then, given the prior $\mu$, for each connected component $C$ of $G$ the ratios encoded by $\f$ determine a probability distribution on $C$. Specifically,
\begin{equation} \label{eq: induced distribution}
    \mu_{\f}(\o'\mid C) = \frac{\mu(\o')\f(\o',\o)}{\sum_{\o''\in C}\mu(\o'')\f(\o'',\o)},
\end{equation}
where $\o$ is any reference state in $C$ (the expression is independent of this choice).
\end{remark}

Similar to the notion of an $F$-cycle and the internal consistency property, we employ the $F$-loop to define external consistency as follows.

\begin{definition} \label{Definition: a loop}
Assume $\rm\bf[INC]$ holds so that $\f$ extends as in \emph{Lemma~\ref{lemma:extension}}, and let $\bigl((\o_i, \bar{\o}_{i})\bigr)_{i=1}^m$  be an $F$-loop.
Then, \emph{External Consistency} $\rm\bf[EXC]$ holds if
\begin{equation}\label{eq: ECON}
   \prod_{i=1}^m \f(\o_i,\bar{\o}_i)=1. 
\end{equation}
\end{definition}

The \textbf{[EXC]} condition has substantive content only when the mediator lacks knowledge of the players’ common knowledge.  Equivalently, if for every $\o$ we have $F(\o) \subseteq C(\o)$, then no $F$-loops arise and \textbf{[EXC]} becomes vacuous. An extreme example is an all-knowing mediator, who knows the exact identity of the realized state, i.e., $F(\o) = \{\o\}$ for every $\o \in \O$. 

We revisit the introductory example to illustrate \textbf{[EXC]}. Consider a target profile assigning $(x,1-x,0,0)$ on $\{\omega_1,\omega_2\}$ and $(0,0,y,1-y)$ on $\{\omega_3,\omega_4\}$, where $x,y\in(0,1)$. By Eq.~\eqref{eq:likelihood}, the signal likelihood ratios needed to induce this profile are
\[
\f(\omega_1,\omega_2) =\frac{x}{1-x}\cdot\frac{\mu(\omega_2)}{\mu(\omega_1)} =\frac{x}{1-x},
\qquad
\f(\omega_4,\omega_3) =\frac{1-y}{y} \cdot \frac{\mu(\omega_3)}{\mu(\omega_4)} =\frac{1-y}{3y}.
\]
Consequently, \textbf{[EXC]} along the loop $((\omega_1,\omega_2),(\omega_4,\omega_3))$ requires $\tfrac{x}{1-x} \cdot \tfrac{1-y}{3y}=1$, which yields $y=\frac{x}{3-2x}$. This is exactly the feasibility restriction derived in the introduction. The factor $1/3$ in the second likelihood ratio comes from the non-uniform prior.

The technical backbone of our analysis, stated in Proposition~\ref{Proposition: basic charecterization}, establishes that internal and external consistency, w.r.t.\ a given PL function~\(\f\), are necessary and sufficient conditions for the existence of an \(F\)-measurable function that reproduces the likelihood ratios encoded by $\f$. (The proof is relegated to the appendix; see Section \ref{Appendix: proof of first main result}.)

\begin{proposition}\label{Proposition: basic charecterization}
Fix $\f$.
There exists an \({F}\)-measurable and strictly positive function \(f\) defined on \(V\) such that, for every \((\omega, \omega') \in E\),
\begin{equation}\label{eq: theorem 1}
\f(\omega, \omega') = \frac{f(\omega)}{f(\omega')},
\end{equation}
if and only if {\rm{\textbf{[INC]}}} and {\rm{\textbf{[EXC]}}} hold for the extension provided by \emph{Lemma~\ref{lemma:extension}}.
\end{proposition}

Once Proposition~\ref{Proposition: basic charecterization} establishes a characterization of a given PL function, we can employ it in Section \ref{Section: Main characterization} to show how one can take a JB, translate it into a PL function, and then replicate this JB for a given mediator.

As stated in Remark~\ref{rmk: phi induces probability}, when $\f$ satisfies \textbf{[INC]} it induces a distribution $\mu_{\f}(\cdot \mid C)$ on each connected component $C$. 
Section~\ref{Section: Main characterization} shows that if $\f$ also satisfies \textbf{[EXC]}, then the mediator can choose an $F$-measurable signaling function $\tau$ and a signal $s$ such that the posterior distribution over $\O$ given $s$ admits the decomposition
\begin{equation} \label{eq: induced distribution on all Omega}
     \mu_{\tau}(\cdot \mid s) = \sum_{C}\mu(C\mid s)\, \mu_{\f}(\cdot \mid C),
\end{equation}
where the sum ranges over all connected components of $G$, and
\[
\mu(C\mid s)=\frac{\sum_{\o \in C} \tau(s\mid \o)\mu(\o)}{\sum_{\o \in \O} \tau(s\mid \o)\mu(\o)}. 
\]
In simple terms, $\mu_{\tau}(\cdot \mid s)$ is the posterior distribution over $\O$ after observing the message $s$, sent by the mediator.

\begin{remark}
    The preceding discussion has focused on the case of public signals, where all players observe the same realization from the information structure. 
    Evidently, if a signaling function induces the joint posteriors of the entire group $N$, then it also induces the joint posteriors of any subgroup $I \subseteq N$. 
    Consequently, \emph{Proposition~\ref{Proposition: basic charecterization}} implies that if the corresponding \emph{PL} function $\f_{\text{JB}}$ satisfies the consistency conditions for $N$ players, it will also do so for any subset of $N$.
\end{remark}

\subsection{Main characterization: implementation to Bayesian updating}\label{Section: Main characterization}

In this section, we apply Proposition~\ref{Proposition: basic charecterization} to the context of Bayesian updating. 
Suppose all players share a common prior $\mu$, and let JB denote a joint belief, as defined in Definition \ref{def: Joint posterior}. 
Can this joint belief be induced by a signaling function $\tau$ of the mediator and a signal $s$?

We employ Proposition \ref{Proposition: basic charecterization} to answer this question. 
To see this, let us focus first on the subset $\O_+ \subseteq \O$ of states $\omega$ for which\footnote{Recall that $\rm{JB}(\omega, i)$ is a distribution while $\rm{JB}(\omega, i)(\o)$ is a probability.} $\rm{JB}(\omega, i)(\o) > 0$ for some player $i$.
Consider the graph $G_+ = (\O_+, E_+)$ , which is the restriction of the original graph $G = (\O, E)$ to the subset $\O_+$. 
Using Eq.~\eqref{eq:likelihood}, define a new PL function $\f_{\text{JB}}$ on the edge set $E_+$ as follows:
\begin{equation}\label{eq: defining phi}
 \f_{\text{JB}}(\omega, \omega') = \frac{{\rm{JB}}(\omega,i)(\o)}{{\rm{JB}}(\omega',i)(\o')} \cdot \frac{\mu(\omega')}{\mu(\omega)}, \ \ \forall \ (\o,\o')\in E_+ \ ,  \o'\in \mathcal{P}_i(\o).
\end{equation}

Since both $\mu$ and $\rm{JB}(\omega, i)(\o)$ are strictly positive on $\Omega_+$, $\f_{\text{JB}}$ is strictly positive as well.
Note that the common realized state constraint (condition (i) in Definition~\ref{def: Joint posterior}), ensures that ${\rm{JB}}(\omega, j)(\o) > 0$ for every player $j$ and every $\omega\in \O_+$. 
Accordingly, the information-set invariance requirement (condition (iii) in Definition~\ref{def: Joint posterior}) ensures that  ${\rm{JB}}(\omega, i) ={\rm{JB}}(\omega', i)$, as long as both are not the zero vector.
Moreover, the likelihood-ratio agreement (condition (iv) in Definition~\ref{def: Joint posterior}) ensures that the right-hand side is invariant to the chosen player $i$, so $\f_{\text{JB}}$ is a well-defined PL function.
Given that ${\rm\textbf{[INC]}}$ holds, we can extend $\f_{\text{JB}}$ to every pair $(\o,\o')$ in a given CKC as done in Lemma \ref{lemma:extension}.
So, whenever ${\rm\textbf{[INC]}}$ holds, we henceforth consider the extended PL function.

Recall that $\boldsymbol{\mu}_{\tau,s}$
(see Eq.\ \eqref{eq: the joint posterior of mu , s} above) denotes the joint posterior associated with $\tau$ and $s$.
As an implication of Proposition~\ref{Proposition: basic charecterization}, the following theorem states that internal and external consistency (w.r.t.\ $\f_{\text{JB}}$) are necessary and sufficient conditions for the existence of an \({F}\)-measurable signaling function $\tau$ and a signal $s$,  such that $\boldsymbol{\mu}_{\tau,s} = \text{JB}$.\footnote{To be clear, the equivalence holds for every player and every state in $\O_+$, specifically, $\forall \o\in\O_+$ and $\forall i\in N$, we have $\mu_{\tau,i}(\cdot|\mathcal{P}_i(\o),s) = \text{JB}(\o,i)$.}
Other than the two consistency conditions, we also require that $\Omega_+$ is $F$-measurable, because any JB that assigns zero mass to some states and positive mass to other states within the same mediator information set is impossible to implement.
(The proof is relegated to the appendix; see Section \ref{Section  - Proof of second theorem}.)

\bigskip 
\begin{theorem}[Main Characterization: Implementable Joint Posteriors]\label{th: main Bayesian}
Fix all agents' partitions, a prior, and a joint belief $\emph{JB}$. 
Then there exists an \({F}\)-measurable signaling function $\tau$ and a signal $s$,  such that $\boldsymbol{\mu}_{\tau,s} = \emph{JB}$ on $\O_+$ and $\{\omega\in\Omega:\tau(s\mid\omega)>0\}=\O_+$ if and only if 
\begin{enumerate}[\emph{(}i\emph{)}]
    \item  $\O_+$ is measurable w.r.t.\ $F$; and
    \item  $\f_{\emph{JB}}$ satisfies conditions {\rm\textbf{[INC]}} and {\rm\textbf{[EXC]}} in $G_+$.\footnote{{\rm\textbf{[EXC]}} is satisfied using the extension provided by Lemma~\ref{lemma:extension}, ensured by {\rm\textbf{[INC]}}.}
\end{enumerate}
\end{theorem}

Note that condition~(i) is necessary because the prior $\mu$ is strictly positive. If the mediator sends a given signal at some $\omega \in \Omega_+$, then the same signal must also be sent at every $\omega' \in F(\omega)$, with the same probability. Consequently, if $\omega' \notin \Omega_+$, the posterior probability assigned to $\omega'$ cannot be zero unless the prior probability of $\omega'$ was already zero.

The characterization also admits a common-prior interpretation. Suppose $\Omega_+$ is nonempty and $F$-measurable. Restrict the information structure to $\Omega_+$ and add an agent whose information partition is the restriction of $F$ and whose beliefs are $\mu(\cdot\mid F(\omega)\cap\Omega_+)$. The target JB is implementable if and only if this augmented belief system admits a strictly positive common prior $\nu$ on $\Omega_+$. The added agent's beliefs ensure that $\nu/\mu$ is $F$-measurable on $\Omega_+$. For sufficiently small $c>0$, the likelihood $\tau(s\mid\omega)=c\,\nu(\omega)/\mu(\omega)$ on $\Omega_+$, with zero likelihood elsewhere, implements the target JB. Conversely, an implementing signal yields such a common prior through
$\nu(\omega)\propto\mu(\omega)\tau(s\mid\omega)$. Our decomposition separates consistency within each common knowledge component from the compatibility of their relative weights across mediator-linked components.

The economic implication of Theorem~\ref{th: main Bayesian} is as follows.  Suppose a function $\f$ is given.  By Remark~\ref{rmk: phi induces probability}, $\f$ induces a probability distribution over each connected component, and thus determines the posterior beliefs of all players.  Theorem~\ref{th: main Bayesian} identifies conditions under which there exists a signaling mechanism, implemented by the mediator, that induces these posterior beliefs, as described in Eq.~\eqref{eq: i'th posterior}.

\section{Non-convexity of the feasible set} \label{Section - non convex set of posteriors}

This section studies the convexity of joint posterior profiles conditional on players' private information.
A convex combination of two profiles uses the same weight at every state and for every player. As the introductory example shows, this operation need not correspond to pooling signals, because Bayesian pooling generally uses different weights at different private information cells.  This non-convexity does not invalidate concavification over public posteriors, that is, posteriors based on the public signal alone. The distinction is between a public posterior and the system of conditional beliefs obtained from it by conditioning on private information.

To understand why convexity may break down, consider the consistency constraints in Theorem~\ref{th: main Bayesian}.  The external consistency condition builds on the function $\f_{\mathrm{JB}}$ defined in Equation~\eqref{eq: defining phi}.  Given either an $F$-loop or an $F$-cycle, take two distinct feasible joint posteriors that individually satisfy the consistency constraints.  For convexity to hold, their average should also satisfy these constraints.  However, the constraints are \emph{nonlinear} in the posterior probabilities, since they arise from a product of likelihood ratios.  So the average may fail to sustain a necessary feasibility condition.

Recall that Definition \ref{def: Joint posterior} allows ${\rm{JB}}(\omega,i)$ to take the value $\emph{\1}_{\emptyset}$, which does not represent a probability distribution. Consequently, the set of joint beliefs is not convex in the standard sense: if one joint belief is a valid distribution and another is $\emph{\1}_{\emptyset}$, their convex combination is not a well-defined belief.  We therefore restrict our attention to cases where, conditional on the state and player, the support of the two joint beliefs are in the same partition elements.  Formally, we adopt the following definition of convexity:
\begin{definition}
    The feasible set is \emph{convex} if for any two feasible joint beliefs ${\rm{JB}}_1$ and ${\rm{JB}}_2$ with matching information-supports \emph{(}i.e., ${\rm{JB}}_1(\omega,i)=\1_{\emptyset} \iff {\rm{JB}}_2(\omega,i)=\1_{\emptyset}$ for all $(\omega,i)$\emph{)}, the candidate posterior profile $\lambda {\rm{JB}}_1 + (1-\lambda){\rm{JB}}_2$ is feasible for all $\lambda\in[0,1]$.\footnote{Matching information-supports does not ensure likelihood-ratio agreement, condition~(iv) of Definition~\ref{def: Joint posterior}. A mixture must satisfy that definition before its induced PL function can be tested for \textbf{[INC]} and \textbf{[EXC]}. A mixture that fails the definition is infeasible.}
\end{definition}

As the introductory example shows, non-convexity can arise even with a single privately informed player.\footnote{\cite{Kolotilin2017} and \cite{Guo2019} provide insights into persuasion problems with privately informed receivers.} We now revisit that example using the consistency conditions.

\begin{example}[The introductory example revisited]\label{ex:  no convexity} \rm
Use $\Omega$, $\mu$, $\mathcal{P}_1$, $F$, and the two-signal experiment $\tau$ from the \hyperref[Section - intro. example]{introductory example}.
Table~\ref{tab: example}, panel A, reports the full posterior profiles induced by $s_1$ and $s_2$; panel B reports their arithmetic average, denoted by $\mathrm{JB}$.
Both signals have strictly positive likelihood at every state, so the two profiles have matching support and their average is a well-defined joint belief.
For the $F$-loop $((\omega_1,\omega_2),(\omega_4,\omega_3))$, the prior-corrected likelihood ratios of this average satisfy
\[
\begin{aligned}
&\frac{\mathrm{JB}(\omega_1,1)(\omega_1)}{\mathrm{JB}(\omega_2,1)(\omega_2)} \cdot
 \frac{\mu(\omega_2)}{\mu(\omega_1)} \cdot 
 \frac{\mathrm{JB}(\omega_4,1)(\omega_4)}{\mathrm{JB}(\omega_3,1)(\omega_3)} \cdot 
 \frac{\mu(\omega_3)}{\mu(\omega_4)}  = \frac{1/2}{1/2}\cdot\frac{3/10}{3/10} \cdot\frac{7/10}{3/10}\cdot\frac{1/10}{3/10} =\frac79\neq 1.
\end{aligned}
\]
Thus, \textbf{[EXC]} fails. By Theorem~\ref{th: main Bayesian}, the average profile is not implementable, and the feasible set is not convex.

\begin{table}[H]
    \centering
    \caption{Signal-specific posteriors and their average}
    \label{tab: example}
    \footnotesize

    \begin{minipage}[t]{0.56\textwidth}
        \centering
        A\par\smallskip
        \begin{tabular}{@{}ccc@{}}
            \toprule
            State
            & $\mathrm{post}(\cdot\mid s_1)$
            & $\mathrm{post}(\cdot\mid s_2)$ \\
            \midrule
            $\o_1$
            & $(\nicefrac{1}{4},\nicefrac{3}{4},0,0)$
            & $(\nicefrac{3}{4},\nicefrac{1}{4},0,0)$ \\
            $\o_2$
            & $(\nicefrac{1}{4},\nicefrac{3}{4},0,0)$
            & $(\nicefrac{3}{4},\nicefrac{1}{4},0,0)$ \\
            $\o_3$
            & $(0,0,\nicefrac{1}{10},\nicefrac{9}{10})$
            & $(0,0,\nicefrac{1}{2},\nicefrac{1}{2})$ \\
            $\o_4$
            & $(0,0,\nicefrac{1}{10},\nicefrac{9}{10})$
            & $(0,0,\nicefrac{1}{2},\nicefrac{1}{2})$ \\
            \bottomrule
        \end{tabular}
    \end{minipage}\hfill
    \begin{minipage}[t]{0.40\textwidth}
        \centering
        B\par\smallskip
        \begin{tabular}{@{}cc@{}}
            \toprule
            State & Average posterior \\
            \midrule
            $\o_1$
            & $(\nicefrac{1}{2},\nicefrac{1}{2},0,0)$ \\
            $\o_2$
            & $(\nicefrac{1}{2},\nicefrac{1}{2},0,0)$ \\
            $\o_3$
            & $(0,0,\nicefrac{3}{10},\nicefrac{7}{10})$ \\
            $\o_4$
            & $(0,0,\nicefrac{3}{10},\nicefrac{7}{10})$ \\
            \bottomrule
        \end{tabular}
    \end{minipage}

    \par\smallskip
    \begin{minipage}{\textwidth}
        \footnotesize
        Note: Panel A reports the player's posterior beliefs given the mediator's signals. Panel B reports their unweighted average,
        $\tfrac12\bigl(
        \mathrm{post}(\cdot\mid s_1)
        +
        \mathrm{post}(\cdot\mid s_2)
        \bigr)$.
    \end{minipage}
\end{table}
\end{example}

Next, one can devise an example with a fully informed mediator and a uniform prior that still does not admit convexity. Fix $\Omega=\{\omega_1,\omega_2,\omega_3,\omega_4\}$, endowed with a uniform prior, a fully informed mediator, while $\mathcal{P}_1=\{\{\omega_1,\omega_2\},\{\omega_3,\omega_4\}\}$ and $\mathcal{P}_2=\{\{\omega_1,\omega_3\},\{\omega_2,\omega_4\}\}$. So there exists a cycle that goes through all four states. Now consider two signals, the first $s_1$ is completely non-informative and the second is $\tau(s_2|\o_1)=\tau(s_2|\o_2)= \tfrac{1}{6}$, $\tau(s_2|\o_3)=\tfrac{2}{6}$ and $\tau(s_2|\o_4)=\tfrac{3}{6}$. One can easily compute the joint posteriors given each signal and their average, and use our characterization to verify that the outcome is a non-implementable joint posterior, although the mediator has full information.

We can also simplify the state space with an example that builds on three states with prior $\mu=(\tfrac14,\tfrac14,\tfrac12)$, where $\mathcal{P}_1=\{\{\omega_1,\omega_2\},\{\omega_3\}\}$, $\mathcal{P}_2=\{\{\omega_1,\omega_3\},\{\omega_2\}\}$, and the mediator's partition is $F=\{\{\omega_1\},\{\omega_2,\omega_3\}\}$. Take two signals where the first is completely non-informative and the second is $\tau(s_2|\o_2)=\tau(s_2|\o_3)=2\tau(s_2|\o_1)$. Again, a simple computation shows that the average of the two joint posterior beliefs is not implementable.

On the other hand, it is easy to show that the feasible set is convex in two benchmark cases of: (i) uninformed players; and (ii) a fully informed mediator with one player.\footnote{Obviously, the set would also be convex in a binary state space, as any connected set is convex in one dimension.} The question now arises: when is the feasible set convex?  Our graph-of-information characterization provides tools to address this question, but it requires two additional notions: \emph{pseudo-independence} and \emph{fully informative loops}.

\begin{definition}
Let $\mathcal{A} = \{A_1, \ldots, A_{\ell}\}$ and 
$\mathcal{B} = \{B_1, \ldots, B_{k}\}$ be two partitions of $\Omega$.   
We say that $\mathcal{A}$ and $\mathcal{B}$ are \emph{pseudo-independent} if, for every strictly positive prior $\mu$, there exist constants 
$a_1, \ldots, a_{\ell}$ and $b_1, \ldots, b_{k}$ such that, for every $i$ and $j$
 with $A_i \cap B_j\not=\emptyset,$
$\mu(A_i \cap B_j) = a_i b_j 
$.
\end{definition}
Technically, pseudo-independence means that, for every prior~$\mu$, 
the incidence matrix
$
\bigl(\mu(A_i \cap B_j)\bigr)_{ij},
$
which represents the joint distribution induced by~$\mu$ over the two partitions, equals the matrix which is the outer product of two vectors,
$(a_1,\ldots,a_{\ell})$ and $(b_1,\ldots,b_{k})$, except for the cells where $A_i \cap B_j=\emptyset.$

This definition establishes a structural condition on the geometry of the partitions. Because the factorization is required to hold for \textit{every} prior, pseudo-independence describes a combinatorial property of the information structures themselves, independent of any specific belief system or probability measure.

\begin{example} \label{ex:pseudo-independet} \rm
Use the same state space and player partition as in Example~\ref{ex:  no convexity}, but take $\mu=(\tfrac15,\tfrac15,\tfrac15,\tfrac25)$. For convenience, denote $A_1 = \{\omega_1, \omega_2\}$ and  $A_2 = \{\omega_3, \omega_4\}$, while the mediator's partition is now $F=\{B_1,B_2,B_3\}$ where \(B_1=\{\omega_1,\omega_3\},B_2 = \{\omega_2\}\), and \(B_3= \{\omega_4\}\).  It is clear that $\mathcal{P}_1$ and $F$ are not probabilistically independent with respect to $\mu$.\footnote{The two partitions are probabilistically independent with respect to $\mu$ if $\mu(A_i \cap B_j) =\mu(A_i) \mu(B_j)$ for every $i, j$.}   However, for the constants $a_1 = a_2 = 1$, $b_1 = b_2 = 0.2$, and $b_3 = 0.4$, we have $\mu(A_i \cap B_j) = a_i b_j $ whenever  $A_i \cap B_j\not = \emptyset$.  For example, $\mu(A_1 \cap B_1) = \mu(\omega_1) = 1 \cdot 0.2 = 0.2$.  Since we can provide a similar factorization for every prior~$\mu$, the partitions are pseudo-independent.  It is important to note that when $A_i \cap B_j = \emptyset$, no condition is imposed.
\end{example}

Next, following \cite{Lagziel2025d}, we define informative and fully informative loops:
\begin{definition} \label{Definition: informative loop}
An F-loop $\bigl((\o_i, \bar{\o}_{i})\bigr)_{i=1}^m$ is \emph{informative} if  there exist an $i$ such that ${\o}_{i}$ and $\bar{\o}_{i}$ belong to two different information sets of $F$. It is \emph{fully informative} if this condition holds for every $i=1,\ldots,m$.
\end{definition}

An informative loop implies that, within a chain of states that players cannot distinguish, the mediator \emph{can} distinguish some states. This capability allows the mediator to induce likelihood ratios that do not aggregate linearly, and may create non-convexity (depending on the prior).

In a single-player setting, the implication of $F(\omega_{i}) \neq F(\bar{\omega}_{i})$ is that $\omega_{i}$ and $\bar{\omega}_{i}$ are indistinguishable to the single player, whereas the mediator can distinguish between them.  In this sense, the mediator may convey additional information to the player along the loop.  For this reason, the loop is called informative. For example, the $F$-loop $((\omega_1,\omega_2),(\omega_4,\omega_3))$ in Example~\ref{ex:  no convexity} is fully informative because $F(\omega_1)\neq F(\omega_2)$ and $F(\omega_4)\neq F(\omega_3)$. Had $F$ been trivial (i.e., $F = \{\Omega\}$), so that the mediator knows nothing, the sequence $((\omega_1,\omega_2)(\omega_4,\omega_3))$ would still form an $F$-loop, but it would no longer be informative.

We are now ready to characterize the convexity of the feasible set. This characterization relies on two conditions. The first is a combinatorial condition on the mediator’s and the player’s partitions, ensuring pseudo-independent factorizations, and the second concerns the information (knowledge) available to the mediator along every loop. (The proof is relegated to the appendix; See Section \ref{Section - proof of convexity theorm}.)

\begin{theorem} \label{Th:convexity}
Consider a player endowed with a partition $\mathcal{P}_1$ and a mediator whose partition is $F$.  The following statements are equivalent.
\begin{itemize}
    \item The feasible set is convex for every prior.
     \item $\mathcal{P}_1$ and $F$ are pseudo-independent.
      \item There is no fully informative $F$-loop.
\end{itemize} 
\end{theorem}

Theorem~\ref{Th:convexity} can be extended to show that, given a fully informative $F$-loop, the feasible set of posterior beliefs is not convex for almost every prior. In other words, a non-convex set of posterior beliefs is generic, in the presence of a fully informative $F$-loop. This result is given in the following Proposition~\ref{prop:generic-nonconvex-informative-loop}. (The proof is relegated to the appendix; See Section \ref{Appendix: proof of generic non-convexity}.)

\begin{proposition}\label{prop:generic-nonconvex-informative-loop}
Consider a player endowed with a partition $\mathcal{P}_1$ and a mediator whose partition is $F$.  If there exists a fully informative $F$-loop, then the set of priors for which the feasible set is convex has Lebesgue measure zero.
\end{proposition}

Interestingly, there are various, quite standard persuasion models with a fully informative $F$-loop.
For example, following \cite{Guo2019}, consider an online platform (Sender) that wants a customer (Receiver) to buy a product. The product has an unknown quality $q \in\{q_H,q_L\}$, where $0<q_L<q_H<1$, known only to the Sender, while the customer reads reviews and forms a private ``type'' $t$: either optimistic $t_H$ (w.p.~$q$) or pessimistic $t_L$ (w.p.~$1-q$). Evidently, the optimistic type is easier to persuade about a high-quality product than the pessimistic one. The Receiver can either buy or not, and conditional on buying, his payoff depends on $q$ (and $0$ otherwise), while the  Sender's payoff is positive only if the Receiver buys (and $0$ otherwise). Thus, the state space comprises $4$ states, where the Receiver's and the Sender's partitions are $\mathcal{P}_1=\{\{q_Ht_H,q_Lt_H\},\{q_Ht_L,q_Lt_L\}\}$ and $F=\{\{q_Ht_H,q_Ht_L\},\{q_Lt_H,q_Lt_L\}\}$, respectively. This yields a $4$-state, fully informative $F$-loop, and the feasible set is non-convex for every strictly positive prior over quality.\footnote{Let $r=\mu(\{q_Ht_H,q_Ht_L\})\in(0,1)$. For the two Receiver cells and the high-quality Sender cell, the prior expression in Eq.~\eqref{eq:knife-edge} equals $r(1-r)(q_H-q_L)>0$. The two feasible profiles constructed in Appendix~\ref{Appendix: proof of generic non-convexity} therefore have an infeasible average.}

\section{Extensions and applications} \label{Section - Extensions and applications}

\subsection{A Mediator as a Blackwell-experiments generator} \label{Section - Generator}

In the previous sections we provided necessary and sufficient conditions for a JB to be generated by a single $F$-measurable signaling function and a single signal. 
Now suppose we are given several beliefs, ${\rm JB}_1,\dots,{\rm JB}_m$. As in Eq.~\eqref{eq: defining phi}, each JB$_i$ induces a PL function $\f_{{\rm JB}_i}$. 
Assume that each $\f_{{\rm JB}_i}$ satisfies the {\textbf{[INC]}} and {\textbf{[EXC]}} conditions, thereby enabling the application of Theorem~\ref{th: main Bayesian}.
This means that for each~$i$, there exists a signaling function~$\tau_i$ and a signal~$s_i$ such that the posterior they induce coincides with~$\mu_i=\mu_{{{\rm JB}_i}}$, as in Eq.~\eqref{eq: induced distribution on all Omega}. Fix these implementations and their induced public posteriors $\mu_1,\ldots,\mu_m$. Thus, each posterior is individually generated by a distinct signaling function and signal. Our question here is whether there exists a single signaling function~$\tau$ that generates all of these posteriors with positive probability, and no others.

To formally introduce this question, denote
\begin{equation} \label{eq: posterior}
    \mu_{\tau}(\omega \mid s) = 
    \frac{\mu(\omega)\tau(s \mid \omega)}{\sum_{\omega' \in \Omega} \mu(\omega') \tau(s \mid \omega')},
\end{equation}
which is the posterior probability of~$\omega$ given that the signal~$s$ has been generated by~$\tau$. 
The corresponding posterior distribution over~$\Omega$ will be denoted by~$\mu_{\tau}(\cdot \mid s)$. 
Note that the signaling function and the signal are constructed so as to retain all posterior probabilities consistent with the respective JBs and subject to the mediator’s informational constraint. 
Once constructed, they induce a distribution over the state space obtained through standard Bayesian updating.

We then ask: under what conditions does there exist a signaling function~$\tau$, measurable with respect to~$F$, such that the signals it produces with positive probability are~$s_1, \ldots, s_m$ and 
\begin{equation} \label{eq: posterior 2}
\bigl\{\mu_{\tau}(\cdot \mid s_1), \ldots, \mu_{\tau}(\cdot \mid s_m)\bigr\}
= \big\{\mu_{_1}, \ldots, \mu_{_m}\bigr\}?
\end{equation}

If the answer is affirmative, the mediator effectively serves as a generator of Blackwell experiments and, in particular, can generate the corresponding posteriors. To investigate this question, we begin with the following definition. A nonzero function \( u : \Omega \to \mathbb{R} \) is called an \emph{option} (or a \emph{state-contingent claim}), as it specifies a monetary payoff for each possible state.

\begin{definition}
\emph{(i)} We say that the family of distributions $\nu_1, \ldots, \nu_k$ \emph{preserves positivity (PP)} w.r.t.\ $\mu$ if, for every option $u$ such that $\mathbb{E}_{\nu_i}[u] \ge 0$ for every $i$, it follows that $\mathbb{E}_{\mu}[u] \ge 0$.
 \\
\emph{(ii)} We say that the family of distributions $\nu_1, \ldots, \nu_k$  \emph{strictly preserves positivity (SPP)}  w.r.t.\ $\mu$ if, for every option $u$ such that $\E_{\mu}(u) = 0$ and $\E_{\nu_i}(u) \ge 0$ for every $i$, it follows that $\E_{\nu_i}(u) = 0$ for every $i$.
\end{definition}

To better understand the notion of positivity preservation, suppose that $\mu$ is the prior belief a decision maker holds about the state space $\Omega$, and let $\mu_1, \dots, \mu_m$ be the posteriors induced by observed signals.  If the posteriors $\mu_1, \dots, \mu_m$ \emph{do not} preserve positivity w.r.t.\ $\mu$, then there exists a utility function assigning payoffs to states such that the expected utility under the prior $\mu$ is strictly lower than the expected utility under each posterior $\mu_i$. This represents a case of time inconsistency: ex ante (i.e., under the prior), the option yields a negative expected reward, while ex post (i.e., conditional on any posterior), the same option yields a positive expected reward, regardless of the specific signal realized.

Strict positivity preservation w.r.t.\ $\mu$ means that the distributions  $\mu_1, \ldots, \mu_k$ are fully aligned with $\mu$ in the following sense:  there is no option whose expected value is nonnegative under all $\mu_i$,  zero under $\mu$, and strictly positive under at least one $\mu_i$.  Formally, if a function $u$ has zero mean under $\mu$, that is, $\mu$ regards it as ``fair'', and each $\mu_i$ weakly favors it (i.e., $\E_{\mu_i}[u] \ge 0$ for all $i$),  then it must be that $\E_{\mu_i}[u] = 0$ for all $i$.  In other words, for any option that is $\mu$-neutral,  if it is weakly favorable for all $\mu_i$, then it is strictly favorable for none.

\begin{obs} 
To show that \emph{SPP} w.r.t.\ $\mu$ implies \emph{PP} w.r.t.\ $\mu$, assume the former.  Consider an option $u$ such that $\mathbb{E}_{\mu_i}[u] \ge 0$ for every $i$ and assume, contrary to  \emph{PP},  that $\mathbb{E}_{\mu}[u] < 0$.  Define the option $v=u-\mathbb{E}_{\mu}[u]$. Then,  $\mathbb{E}_{\mu_i}[v] =\mathbb{E}_{\mu_i}[u]-\mathbb{E}_{\mu}[u] \ge 0$ and  $\mathbb{E}_{\mu}[v] =\mathbb{E}_{\mu}[u]-\mathbb{E}_{\mu}[u] = 0$.  By \emph{SPP}, $\mathbb{E}_{\mu_i}[v] =0$  and thus $\mathbb{E}_{\mu_i}[u]= \mathbb{E}_{\mu}[u] < 0$ for every $i$, which contradict the assumption. 
\end{obs}

We can now discuss the following Theorem \ref{Theorem: multiple JP} that extends previous results to a set of joint posteriors. Retain the public posteriors $\mu_1,\ldots,\mu_m$ from the fixed implementations chosen above.\footnote{For the JBs themselves, the public representatives may be chosen. There exists a single $F$-measurable experiment generating every $\mathrm{JB}_i$ with positive probability and no additional JB profiles if and only if one can choose, for each $i$, a public posterior $\nu_i$ induced by an $F$-measurable implementation of $\mathrm{JB}_i$ such that the family $\{\nu_1,\ldots,\nu_m\}$ satisfies SPP w.r.t.\ $\mu$. Sufficiency follows from Theorem~\ref{Theorem: multiple JP}. For necessity, pool all signals inducing the same target JB. Each pooled signal still implements that JB, and the resulting public representatives average to the prior with strictly positive weights, hence satisfy SPP.} The first part of the following Theorem~\ref{Theorem: multiple JP} states that PP is a necessary and sufficient condition for the existence of a strategy $\tau$ whose signals generate posterior distributions, all of which are contained in \(\{\mu_{1}, \ldots, \mu_{m}\}\). 
The second part of the theorem uses SPP to characterize when there is a signaling function whose set of posterior coincides with \(\{\mu_{1}, \ldots, \mu_{m}\}\).
(The proof is given in Section \ref{Appendix: proof of third theorem} in the appendix.)

\begin{theorem}\label{Theorem: multiple JP} Let \(\mu\) be a common prior, and let \(\f_{{\rm JB}_1}, \ldots, \f_{{\rm JB}_m}\)  be \emph{PL} functions that satisfy \emph{\textbf{[INC]}}  and \emph{\textbf{[EXC]}}. 
For every ${\rm JB}_i$, assume that $\Omega_+^i$, defined as in  \emph{Theorem \ref{th: main Bayesian}}, is measurable w.r.t. $F$. For each~$i$, fix a public posterior $\mu_i$ induced by an $F$-measurable implementation of $\mathrm{JB}_i$. Then, there exists an \({F}\)-measurable signaling function \(\tau\) such that: \\
(i) for every signal \(s\) generated by \(\tau\) with positive probability, the posterior it induces belongs to the family \(\{\mu_{1}, \ldots, \mu_{m}\}\) if and only if this family \emph{PP} w.r.t.\  \(\mu\);\\
(ii) the set of posteriors it generates with positive probability coincides with \(\{\mu_{1}, \ldots, \mu_m\}\)  if and only if this family \emph{SPP} w.r.t.\  \(\mu\).
\end{theorem}

One should note the construction underlying Theorem \ref{Theorem: multiple JP}.  
First, a common prior and a set of JBs are fixed to satisfy the conditions of Theorem \ref{th: main Bayesian}.  
Next, the implied strategies are employed to construct a set of posteriors over $\Omega$ that abstract from the players’ private information.  
The necessary and sufficient conditions stated in Theorem \ref{Theorem: multiple JP} are then imposed on these posteriors, relative to the prior, to guarantee the existence of an $F$-measurable unified strategy.  
Once this strategy is applied, together with the common prior and the players’ private information, it induces the desired JBs.

\subsection{Fact-checking and depolarization procedures}

The issue of polarization, either ideological or affective, has gained much concern in public
discourse, and therefore attention both in general and in research.\footnote{See, for instance,
\cite{Nimark2019,Matejka2021,Levy2021,Yuksel2022,Perego2022,Bowen2023,Ikan2025,Goldstein2025}
among many others.}
Yet, there has been considerably less focus on methods of \textit{reducing} polarization that naturally align with our framework, and can provide insights into this issue.

For that purpose, we interpret the mediator as a fact-checker (or some other verification /
regulatory authority) operating on a news or social-media platform. The underlying state
$\omega\in\Omega$ describes both the true content of a story and the relevant context. Users are
partitioned into groups $i\in N$ with information partitions $\mathcal{P}_i$ (``echo chambers''), and
the fact-checker has a partition $F$ capturing the evidence it can access (e.g., databases, legal
documents). Accordingly, the fact-checking policy is an $F$-measurable public experiment
$\tau:\Omega\to\Delta(S)$ whose realizations $s\in S$ are labels such as ``True'',
``Misleading'', or ``False''.

Given $p\in[0,1]$, we say that a strategy $\tau$ induces \emph{$p$--consensus} if
$$
D(\tau) = \E_{\mu,\tau} \left[\max_{i,j\in N} \|\bigl(\mu_{\tau,i}(\cdot \mid \mathcal{P}_i(\omega),s)\bigr) - \bigl(\mu_{\tau,j}(\cdot \mid \mathcal{P}_j(\omega),s)\bigr) \|\right] \leq 2(1-p),
$$
where $\|\cdot \|$ is the $l_1$-norm and $D(\tau)$ is referred to as the \emph{expected disagreement} given $\tau$.
Note that the case where players have a common prior, and no private information relates to $p=1$, a situation we refer to as \emph{a full consensus}.

Evidently, the case of full consensus is very demanding.
It implies that all disagreements completely vanish, and intuitively, this would require an overwhelming amount of information on the side of the mediator.
The following proposition captures this. 
(The proof is straightforward, thus omitted.)

\begin{proposition} \label{prop:full-consensus}
The mediator can induce full consensus for every common prior if and only if
$\mathcal{P}_i\vee F =\mathcal{P}_j\vee F$ for every $i,j\in N$.\footnote{$\mathcal{P}_i\vee F$ is the coarsest common refinement of $\mathcal{P}_i$ and $F$, i.e., the partition generated by intersecting their cells.}
\end{proposition}

The characterizing criterion is quite restrictive:  it requires the mediator to bridge \emph{any} informational gap between any two individuals.
In many environments this is unrealistic.
A potentially more reasonable and modest objective is to reduce polarization without necessarily eliminating it completely, that is, achieving $p$-consensus for some $p<1$.
In this case, our analysis and results become much relevant.

The mediator’s problem in this section can be interpreted as choosing an $F$-measurable experiment $\tau$ to maximize consensus, or equivalently to minimize $D(\tau)$. Theorem~\ref{th: main Bayesian} characterizes the joint posterior profiles that can arise after each signal. The objective $D(\tau)$ also depends on the probabilities of these signals, so optimizing it is not the same as optimizing over individual feasible profiles.

The next example illustrates that, under this objective, full revelation of the mediator’s partition $F$ is generally \emph{not} optimal: a strictly less informative, noisy experiment can yield strictly lower expected disagreement.

\begin{example}  \label{ex:polarization-noisy-better} \rm 
Consider a probability space with four states $\Omega=\{\omega_1,\omega_2,\omega_3,\omega_4\}$ with the prior $\mu=\bigl(\tfrac{1}{16},\tfrac{4}{16},\tfrac{5}{16},\tfrac{6}{16} \bigr)$.
There are two players and a mediator whose partitions are given by $\mathcal{P}_1=\bigl\{\{\omega_1,\omega_3,\omega_4\},\ \{\omega_2\}\bigr\}$,  $\mathcal{P}_2=\bigl\{\{\omega_1\},\ \{\omega_2,\omega_3,\omega_4\}\bigr\}$, and $F=\bigl\{\{\omega_1,\omega_3\},\ \{\omega_2,\omega_4\}\bigr\}$, respectively.

Given that the mediator provides no information (a constant signal), each player simply conditions on $\mathcal{P}_i(\omega)$, and the resulting expected disagreement is approximately $0.848$.
However, under a fully revealing policy, the expected disagreement drops to $0.808$, so full revelation of $F$ is helpful in reducing polarization relative to having no mediator.
But the mediator could even do better.
For example, consider a strictly less informative  $\tau^{\mathrm{noisy}}$
with two signals $s_1,s_2$:
\[
\tau^{\mathrm{noisy}}(s_1\mid\omega_1)
=
\tau^{\mathrm{noisy}}(s_1\mid\omega_3)
=0,
\qquad
\tau^{\mathrm{noisy}}(s_1\mid\omega_2)
=
\tau^{\mathrm{noisy}}(s_1\mid\omega_4)
=\tfrac34,
\]
and $\tau^{\mathrm{noisy}}(s_2\mid\omega)=1-\tau^{\mathrm{noisy}}(s_1\mid\omega)$ for all $\omega$.
This experiment partially pools the two $F$-cells and is therefore strictly less informative (in the Blackwell sense) than the completely informative strategy.
A direct calculation shows that the expected disagreement under $\tau^{\mathrm{noisy}}$ is $0.775$.

Thus, although full revelation of $F$ reduces polarization relative to having no mediator, it is \emph{strictly dominated} (for the purpose of minimizing expected disagreement) by a coarser, noisy $F$-measurable experiment.  Evidently, the reason is the interaction between a fully revealing $F$ and the  partitions $(\mathcal{P}_i)_i$, through the consistency constraints in Theorem \ref{th: main Bayesian}. This interaction creates states at which players’ posteriors diverge sharply.  By partially pooling $F$-cells, the mediator can smooth these differences and achieve a higher agreement level.
\end{example}

Example~\ref{ex:polarization-noisy-better} shows that once we move from full consensus to a more reasonable objective, the mediator faces a genuine information-design problem. The graph-of-information structure and the consistency conditions dictate the feasible joint posteriors, but within this feasible set, it is generally \emph{not} optimal to reveal all available information, as garblings of $F$ can still reduce polarization more effectively.

\subsection{A general framework for information design and persuasion}
 
Our study provides a general information framework that could be applied in various settings within the realm of information design. For example, in a benchmark binary state-space setting, where players have no private information and with a fully informed mediator, Theorems \ref{th: main Bayesian} and \ref{Theorem: multiple JP} collapse to the standard splitting lemma: Bayes plausibility is the only feasibility constraint.  This recovers the classical concavification approach of \cite{Aumann1995} and \cite{Kamenica2011} as a special case of our characterization.

In addition, our framework relates to the information-design questions studied in the optimal disclosure in auctions literature. In models such as \cite{Eso2007}, including the handicap auction, a seller controls the release of additional private signals about bidders' valuations without observing their realizations, and integrates disclosure with auction design. Likewise, in \cite{Bergemann2007}, the seller's disclosure choice is modeled as bidder-specific information structures whose realized signals are privately observed by bidders. Within finite-state public-disclosure environments with fixed initial information partitions, when the seller does not have perfect information or bidders also hold initial private signals (as in more general information-design problems with privately informed agents, as in \citealp{Candogan2023}), our internal and external consistency conditions describe exactly which belief patterns about values can be generated by a feasible public disclosure policy.

In our language, the underlying state is a realized $\omega$ (e.g., product quality or the bidders' value profile), each bidder's ex ante information is a fixed partition $\mathcal{P}_i$, and the seller/auctioneer is exactly a mediator whose knowledge is described by a partition $F$. The public disclosure rule is an $F$-measurable Blackwell experiment, so for any finite-state auction environment with fixed initial information partitions and public disclosure, the induced beliefs across bidders and states form a joint posterior to which Theorems \ref{th: main Bayesian} and \ref{Theorem: multiple JP} apply.

When the mediator is assigned an objective, it is important to distinguish public posteriors from the joint posterior profiles obtained after conditioning on private information. Section~\ref{Section - How to optimize} discusses the mediator's optimization problem once an objective is specified.

\subsection{Mediator's optimization} \label{Section - How to optimize}

To address the question of mediator optimization, one should first specify the objective function of the mediator.  For example, the mediator may seek to maximize the probability that a particular action is chosen, or, in a strategic application, a weighted average of the players' equilibrium expected payoffs.

In a strategic application, one must also consider the issue of multiple equilibria.
Upon observing the mediator's signal, players update their beliefs and then play an incomplete-information game.  If this game has multiple equilibria, evaluating the mediator's objective requires an equilibrium-selection rule.

Once the objective function and any required equilibrium-selection rule are fixed, the mediator chooses an $F$-measurable experiment. Every joint posterior profile it generates must be feasible. The introductory example shows that a convex combination of feasible profiles need not satisfy this requirement.

\heading{Optimal trade in the introductory example.}
Return to the introductory technology-adoption example. Let $z=\Pr(q=1\mid s)$ be the public posterior probability of good quality. Because the provider's report depends only on quality, the firm's conditional beliefs and the probability of a favorable assessment are
\[
x(z)=\frac{3z}{2+z},\qquad
y(z)=\frac{z}{2-z},\qquad
\Pr(\text{favorable assessment}\mid s)=\frac{2+z}{4}.
\]
Given the stated adoption cost $k=\tfrac12$, a firm with a favorable assessment adopts exactly when $z\ge\tfrac25$, and one with an unfavorable assessment adopts exactly when $z\ge\tfrac23$. Normalizing the provider's benefit from adoption to one, its auxiliary payoff, in the sense of \citet[Section~VI.A]{Kamenica2011}, is given by
\[
V(z)=
\begin{cases}
0, & 0\le z<\tfrac25,\\
\tfrac{2+z}{4}, & \tfrac25\le z<\tfrac23,\\
1, & \tfrac23\le z\le1.
\end{cases}
\]
This function accounts for the firm's private assessment by weighting its adoption decisions across the two information cells. Since $V(z)\le\tfrac32z$ and Bayes plausibility gives $\E[z]=\tfrac25$, every feasible experiment satisfies $\Pr(\text{trade})=\E[V(z)] \le\tfrac32\cdot\tfrac25=\tfrac35$. The bound is attained by a binary experiment whose adoption report has state-wise likelihoods $(1,\tfrac13,1,\tfrac13)$. This report occurs with probability $\tfrac35$ and induces $(x,y)=(\tfrac34,\tfrac12)$, so both types adopt. Its complementary signal reveals bad quality, resulting in no trade.

If the provider also observes the private assessment, its information partition consists of singletons. Under any disclosure policy, adoption requires an expected gross return of at least $k$, so $\tfrac12\Pr(\text{trade}) \le\Pr(q=1)=\tfrac25$. The resulting upper bound $\Pr(\text{trade})\le\tfrac45$ is attained by an adoption report with likelihoods $(1,1,1,\tfrac13)$ and its complementary report. The adoption report induces $(x,y)=(\tfrac12,\tfrac12)$ and occurs with probability $\tfrac45$; the complementary report reveals bad quality. Thus, at the same adoption cost, the minimum probability of no trade is $\tfrac25$ when the provider observes only quality, compared with $\tfrac15$ when it also observes the assessment: informational constraints double the minimum probability of no trade from $20\%$ to $40\%$.

\subsection{An additive interpretation and implications to potential games} \label{Section - implications to potential games} The graph-theoretic constraints derived for Bayesian updating reflect a structural property applicable beyond information design. By applying a logarithmic transformation to the internal consistency condition, we obtain an additive constraint that characterizes \emph{exact potential games} \citep{monderer1996potential}.

Consider a directed graph $G=(V, E)$ containing the reverse of every edge, with edge weights $\rho: E \to \mathbb{R}$. We define \emph{Additive Internal Consistency} \textrm{\textbf{[INC-ADD]}} as the requirement that for every directed cycle (including two-edge returns) $(\omega_1, \dots, \omega_{m+1}=\omega_1)$:
\begin{equation} \label{eq: additive inc}
\sum_{i=1}^m \rho(\omega_i, \omega_{i+1}) = 0.
\end{equation}
Taking $\f=e^\rho$ and $F$ to be the partition into singletons, Proposition~\ref{Proposition: basic charecterization} and the logarithmic transformation $P=-\log f$ imply that \textrm{\textbf{[INC-ADD]}} is necessary and sufficient for the existence of a potential function $P:V\to\mathbb R$ such that $\rho(\omega,\omega')=P(\omega')-P(\omega)$. 

This structure applies directly to strategic games. Let the vertices of the \emph{game graph} be action profiles, with directed edges representing unilateral deviations. In this context, a cycle corresponds to a sequence of unilateral deviations that returns to the initial action profile.

We weight each deviation by the deviator's utility difference:
\[
\rho\left(a, (a_i', a_{-i})\right) := u_i(a_i', a_{-i}) - u_i(a).
\]

\begin{corollary} \label{cor: 2}
A strategic game $\G$ is an exact potential game if and only if the associated edge function $\rho$ satisfies \textrm{\textbf{[INC-ADD]}} on every cycle of the game graph.
\end{corollary}

This observation unifies the two frameworks. In our original setting, the mediator constructs a global belief system consistent with local likelihood ratios, while a potential game implies a global potential whose changes equal the deviator's utility changes. In both frameworks, the obstruction to ``globalizing'' local information is the non-vanishing of accumulated weights along cycles in the underlying graph.

\section{Conclusion}
This paper analyzes the implementation of joint posterior beliefs in environments where agents possess private information and a non-strategic mediator holds partial information. By modeling the informational constraints of the mediator as measurability restrictions with respect to a specific partition, we provide a necessary and sufficient characterization for the rationalization of joint beliefs.

The core theoretical contribution lies in the identification of two structural conditions: \textit{Internal Consistency} (INC) and \textit{External Consistency} (EXC).  The study demonstrates that a joint posterior with $F$-measurable support is implementable if and only if the product of prior-corrected likelihood ratios along every $F$-cycle and $F$-loop in the graph restricted to that support equals one. This characterization effectively translates abstract measurability constraints into verifiable algebraic properties of the posterior likelihood function.

A significant implication of this analysis is the finding that the set of feasible joint posteriors is generally \emph{non-convex}. In the baseline Bayesian persuasion model of \cite{Kamenica2011}, Bayes plausibility characterizes implementable distributions of posteriors; here, informational constraints alone can make the set of posterior profiles conditional on private information non-convex. We further characterize, in the single-player setting, the conditions of pseudo-independence and the absence of fully informative loops under which convexity holds for every strictly positive prior.

Finally, the paper extends these insights to broader economic applications. It generalizes the framework to the generation of multiple posteriors via the \emph{preservation of positivity} condition. Furthermore, the developed graph-theoretic approach is shown to be relevant for diverse fields, offering a connection to potential games through additive internal consistency and providing insights into belief depolarization, where coarser information structures may outperform full revelation in minimizing disagreement.

\bibliographystyle{ecta}
\bibliography{MyCollection}

\begin{thebibliography}{43}
\newcommand{\enquote}[1]{``#1''}
\expandafter\ifx\csname natexlab\endcsname\relax\def\natexlab#1{#1}\fi

\bibitem[\protect\citeauthoryear{Arieli, Babichenko, Sandomirskiy, and Tamuz}{Arieli et~al.}{2021}]{Arieli2021}
\textsc{Arieli, I., Y.~Babichenko, F.~Sandomirskiy, and O.~Tamuz} (2021): \enquote{Feasible Joint Posterior Beliefs,} \emph{Journal of Political Economy}, 129, 2546--2594.

\bibitem[\protect\citeauthoryear{Aumann}{Aumann}{1976}]{Aumann1976}
\textsc{Aumann, R.~J.} (1976): \enquote{Agreeing to {Disagree},} \emph{The Annals of Statistics}, 4, 1236--1239.

\bibitem[\protect\citeauthoryear{Aumann and Maschler}{Aumann and Maschler}{1995}]{Aumann1995}
\textsc{Aumann, R.~J. and M.~Maschler} (1995): \emph{Repeated {Games} with {Incomplete} {Information}}, Cambridge, MA, USA: MIT Press.

\bibitem[\protect\citeauthoryear{Ball and Esp{\'i}n-S{\'a}nchez}{Ball and Esp{\'i}n-S{\'a}nchez}{2021}]{Ball2021}
\textsc{Ball, I. and J.-A. Esp{\'i}n-S{\'a}nchez} (2021): \enquote{Experimental {Persuasion},} {SSRN} working paper 3906869.

\bibitem[\protect\citeauthoryear{Bergemann and Pesendorfer}{Bergemann and Pesendorfer}{2007}]{Bergemann2007}
\textsc{Bergemann, D. and M.~Pesendorfer} (2007): \enquote{Information structures in optimal auctions,} \emph{Journal of Economic Theory}, 137, 580--609.

\bibitem[\protect\citeauthoryear{Blackwell}{Blackwell}{1951}]{blackwell1951}
\textsc{Blackwell, D.} (1951): \enquote{{Comparison of Experiments},} in \emph{Proceedings of the Second Berkeley Symposium on Mathematical Statistics and Probability}, Berkeley, Calif.: University of California Press, 93--102.

\bibitem[\protect\citeauthoryear{Bowen, Dmitriev, and Galperti}{Bowen et~al.}{2023}]{Bowen2023}
\textsc{Bowen, T.~R., D.~Dmitriev, and S.~Galperti} (2023): \enquote{Learning from {Shared} {News}: {When} {Abundant} {Information} {Leads} to {Belief} {Polarization},} \emph{The Quarterly Journal of Economics}, 138, 955--1000.

\bibitem[\protect\citeauthoryear{Burdzy and Pitman}{Burdzy and Pitman}{2020}]{Burdzy2020}
\textsc{Burdzy, K. and J.~Pitman} (2020): \enquote{Bounds on the probability of radically different opinions,} \emph{Electronic Communications in Probability}, 25, 1--12, paper No. 14.

\bibitem[\protect\citeauthoryear{Candogan and Strack}{Candogan and Strack}{2023}]{Candogan2023}
\textsc{Candogan, O. and P.~Strack} (2023): \enquote{Optimal disclosure of information to privately informed agents,} \emph{Theoretical Economics}, 18, 1225--1269.

\bibitem[\protect\citeauthoryear{Chen, Lin, Tang, and Tucker-Foltz}{Chen et~al.}{2025}]{Chen2025}
\textsc{Chen, Y., T.~Lin, W.~Tang, and J.~Tucker-Foltz} (2025): \enquote{Explainable {Information} {Design},} ArXiv:2508.14196 [cs].

\bibitem[\protect\citeauthoryear{Dawid, DeGroot, Mortera, Cooke, French, Genest, Schervish, Lindley, McConway, and Winkler}{Dawid et~al.}{1995}]{Dawid1995}
\textsc{Dawid, A.~P., M.~H. DeGroot, J.~Mortera, R.~Cooke, S.~French, C.~Genest, M.~J. Schervish, D.~V. Lindley, K.~J. McConway, and R.~L. Winkler} (1995): \enquote{Coherent combination of experts' opinions,} \emph{Test}, 4, 263--313.

\bibitem[\protect\citeauthoryear{Es{\H o} and Szentes}{Es{\H o} and Szentes}{2007}]{Eso2007}
\textsc{Es{\H o}, P. and B.~Szentes} (2007): \enquote{Optimal {Information} {Disclosure} in {Auctions} and the {Handicap} {Auction},} \emph{The Review of Economic Studies}, 74, 705--731.

\bibitem[\protect\citeauthoryear{Feinberg}{Feinberg}{2000}]{feinberg2000characterizing}
\textsc{Feinberg, Y.} (2000): \enquote{Characterizing Common Priors in the Form of Posteriors,} \emph{Journal of Economic Theory}, 91, 127--179.

\bibitem[\protect\citeauthoryear{Geanakoplos}{Geanakoplos}{1994}]{Geanakoplos1994}
\textsc{Geanakoplos, J.} (1994): \enquote{{Common} knowledge,} in \emph{Handbook of {Game} {Theory} with {Economic} {Applications}}, ed. by R.~J. Aumann and S.~Hart, Elsevier, vol.~2, chap.~40, 1437--1496.

\bibitem[\protect\citeauthoryear{Goldstein, Lagziel, and Raveh}{Goldstein et~al.}{2025}]{Goldstein2025}
\textsc{Goldstein, N., D.~Lagziel, and O.~Raveh} (2025): \enquote{Political {Rational} {Inattention}: {A} {New} {Measure} {With} an {Application} to {Political} {Polarization},} {SSRN} working paper 5210838.

\bibitem[\protect\citeauthoryear{Guo and Shmaya}{Guo and Shmaya}{2019}]{Guo2019}
\textsc{Guo, Y. and E.~Shmaya} (2019): \enquote{The Interval Structure of Optimal Disclosure,} \emph{Econometrica}, 87, 653--675.

\bibitem[\protect\citeauthoryear{Gutmann, Kemperman, Reeds, and Shepp}{Gutmann et~al.}{1991}]{Gutmann1991}
\textsc{Gutmann, S., J.~H.~B. Kemperman, J.~A. Reeds, and L.~A. Shepp} (1991): \enquote{Existence of {Probability} {Measures} with {Given} {Marginals},} \emph{The Annals of Probability}, 19, 1781--1797.

\bibitem[\protect\citeauthoryear{Harsanyi}{Harsanyi}{1967--1968}]{harsanyi1967games}
\textsc{Harsanyi, J.~C.} (1967--1968): \enquote{Games with Incomplete Information Played by 'Bayesian' Players, Parts I–III,} \emph{Management Science}, 14, 159--182, 320--334, and 486--502, part I: The Basic Model; Part II: Bayesian Equilibrium Points; Part III: The Basic Probability Distribution of the Game.

\bibitem[\protect\citeauthoryear{Heifetz}{Heifetz}{2006}]{heifetz2006positive}
\textsc{Heifetz, A.} (2006): \enquote{The Positive Foundation of the Common Prior Assumption,} \emph{Games and Economic Behavior}, 56, 105--120.

\bibitem[\protect\citeauthoryear{Hellman and Samet}{Hellman and Samet}{2012}]{Hellman2012}
\textsc{Hellman, Z. and D.~Samet} (2012): \enquote{How common are common priors?} \emph{Games and Economic Behavior}, 74, 517--525.

\bibitem[\protect\citeauthoryear{Hellwig}{Hellwig}{2013}]{Hellwig2013}
\textsc{Hellwig, M.~F.} (2013): \enquote{From posteriors to priors via cycles: {An} addendum,} \emph{Economics Letters}, 118, 455--458.

\bibitem[\protect\citeauthoryear{Herings, Karos, and Kerman}{Herings et~al.}{2024}]{Herings2024}
\textsc{Herings, P. J.-J., D.~Karos, and T.~T. Kerman} (2024): \enquote{Belief inducibility and informativeness,} \emph{Theory and Decision}, 96, 517--553.

\bibitem[\protect\citeauthoryear{Ikan, Lagziel, and Raveh}{Ikan et~al.}{2025}]{Ikan2025}
\textsc{Ikan, L., D.~Lagziel, and O.~Raveh} (2025): \enquote{Resource windfalls, connectivity, and political polarization,} \emph{Journal of Environmental Economics and Management}, 132, 103164.

\bibitem[\protect\citeauthoryear{Kamenica and Gentzkow}{Kamenica and Gentzkow}{2011}]{Kamenica2011}
\textsc{Kamenica, E. and M.~Gentzkow} (2011): \enquote{{Bayesian Persuasion},} \emph{American Economic Review}, 101, 2590--2615.

\bibitem[\protect\citeauthoryear{Kellerer}{Kellerer}{1961}]{Kellerer1961}
\textsc{Kellerer, H.~G.} (1961): \enquote{Funktionen auf {Produkträumen} mit vorgegebenen {Marginal}-{Funktionen},} \emph{Mathematische Annalen}, 144, 323--344.

\bibitem[\protect\citeauthoryear{Kolotilin, Mylovanov, Zapechelnyuk, and Li}{Kolotilin et~al.}{2017}]{Kolotilin2017}
\textsc{Kolotilin, A., T.~Mylovanov, A.~Zapechelnyuk, and M.~Li} (2017): \enquote{Persuasion of a {Privately} {Informed} {Receiver},} \emph{Econometrica}, 85, 1949--1964.

\bibitem[\protect\citeauthoryear{Kosenko}{Kosenko}{2020}]{Kosenko2020}
\textsc{Kosenko, A.} (2020): \enquote{Mediated {Persuasion},} ArXiv:2012.00098 [econ].

\bibitem[\protect\citeauthoryear{Lagziel, Lehrer, and Wang}{Lagziel et~al.}{2025}]{Lagziel2025d}
\textsc{Lagziel, D., E.~Lehrer, and T.~Wang} (2025): \enquote{Comparison of {Oracles}: {Part} {I},} ArXiv:2505.15955 [econ].

\bibitem[\protect\citeauthoryear{Lagziel, Lehrer, and Wang}{Lagziel et~al.}{2026}]{Lagziel2026a}
---\hspace{-.1pt}---\hspace{-.1pt}--- (2026): \enquote{Comparison of {Deterministic} {Information} {Providers},} ArXiv:2609.18701 [econ.TH].

\bibitem[\protect\citeauthoryear{Lehrer and Samet}{Lehrer and Samet}{2014}]{lehrer2014belief}
\textsc{Lehrer, E. and D.~Samet} (2014): \enquote{Belief Consistency and Trade Consistency,} \emph{Games and Economic Behavior}, 83, 165--177.

\bibitem[\protect\citeauthoryear{Levy}{Levy}{2021}]{Levy2021}
\textsc{Levy, R.} (2021): \enquote{Social {Media}, {News} {Consumption}, and {Polarization}: {Evidence} from a {Field} {Experiment},} \emph{American Economic Review}, 111, 831--870.

\bibitem[\protect\citeauthoryear{Mat{\v e}jka and Tabellini}{Mat{\v e}jka and Tabellini}{2021}]{Matejka2021}
\textsc{Mat{\v e}jka, F. and G.~Tabellini} (2021): \enquote{Electoral {Competition} with {Rationally} {Inattentive} {Voters},} \emph{Journal of the European Economic Association}, 19, 1899--1935.

\bibitem[\protect\citeauthoryear{Monderer and Shapley}{Monderer and Shapley}{1996}]{monderer1996potential}
\textsc{Monderer, D. and L.~S. Shapley} (1996): \enquote{Potential games,} \emph{Games and Economic Behavior}, 14, 124--143.

\bibitem[\protect\citeauthoryear{Morris}{Morris}{1994}]{morris1994trade}
\textsc{Morris, S.} (1994): \enquote{Trade with Heterogeneous Prior Beliefs and Asymmetric Information,} \emph{Econometrica}, 62, 1327--1347.

\bibitem[\protect\citeauthoryear{Morris}{Morris}{2020}]{Morris2020}
---\hspace{-.1pt}---\hspace{-.1pt}--- (2020): \enquote{No {Trade} and {Feasible} {Joint} {Posterior} {Beliefs},} Working paper, Massachusetts Institute of Technology.

\bibitem[\protect\citeauthoryear{Nimark and Sundaresan}{Nimark and Sundaresan}{2019}]{Nimark2019}
\textsc{Nimark, K.~P. and S.~Sundaresan} (2019): \enquote{Inattention and belief polarization,} \emph{Journal of Economic Theory}, 180, 203--228.

\bibitem[\protect\citeauthoryear{Perego and Yuksel}{Perego and Yuksel}{2022}]{Perego2022}
\textsc{Perego, J. and S.~Yuksel} (2022): \enquote{Media {Competition} and {Social} {Disagreement},} \emph{Econometrica}, 90, 223--265.

\bibitem[\protect\citeauthoryear{Rockafellar}{Rockafellar}{1970}]{rockafellar1970convex}
\textsc{Rockafellar, R.~T.} (1970): \emph{Convex Analysis}, Princeton, NJ: Princeton University Press.

\bibitem[\protect\citeauthoryear{Rodrigues-Neto}{Rodrigues-Neto}{2009}]{Rodrigues-Neto2009}
\textsc{Rodrigues-Neto, J.~A.} (2009): \enquote{From posteriors to priors via cycles,} \emph{Journal of Economic Theory}, 144, 876--883.

\bibitem[\protect\citeauthoryear{Strassen}{Strassen}{1965}]{Strassen1965}
\textsc{Strassen, V.} (1965): \enquote{The {Existence} of {Probability} {Measures} with {Given} {Marginals},} \emph{The Annals of Mathematical Statistics}, 36, 423--439.

\bibitem[\protect\citeauthoryear{Yuksel}{Yuksel}{2022}]{Yuksel2022}
\textsc{Yuksel, S.} (2022): \enquote{Specialized {Learning} and {Political} {Polarization},} \emph{International Economic Review}, 63, 457--474.

\bibitem[\protect\citeauthoryear{Zaslavsky}{Zaslavsky}{1989}]{Zaslavsky1989}
\textsc{Zaslavsky, T.} (1989): \enquote{Biased graphs. {I}. {Bias}, balance, and gains,} \emph{Journal of Combinatorial Theory, Series B}, 47, 32--52.

\bibitem[\protect\citeauthoryear{Ziegler}{Ziegler}{2020}]{Ziegler2020}
\textsc{Ziegler, G.} (2020): \enquote{Adversarial bilateral information design,} Working paper.

\end{thebibliography}

\appendix 

\section{Proofs}\label{sec: appendix proofs}

\subsection{Proof of Lemma \ref{lemma:extension} } \label{Appendix: proof of extension Lemma}

For every pair \((\omega_1, \omega_{m+1})\) such that \(\omega_1 \twoheadrightarrow \omega_{m+1}\), define
\[
\f(\omega_1, \omega_{m+1}) := \prod_{i=1}^m \f(\omega_i, \omega_{i+1}),
\]
provided that the sequence \(\bigl((\omega_i, \omega_{i+1})\bigr)_{i=1}^m\) lies in \(E\). To show that this is well defined, suppose there are two sequences connecting \(\omega_1\) and \(\omega_{m+1}\): 
\[
\bigl((\omega_i, \omega_{i+1})\bigr)_{i=1}^m \quad \text{and} \quad \bigl((\xi_j, \xi_{j+1})\bigr)_{j=1}^k,
\]
with \(\xi_1 = \omega_1\) and \(\xi_{k+1} = \omega_{m+1}\). 
Concatenating the two sequences, with the first in its original order and the second reversed, produces an F-cycle.

By \textbf{[INC]} and Eq.~(\ref{eq:phi_reversed}), we have
\[\prod_{i=1}^m \f(\omega_i, \omega_{i+1}) \cdot \prod_{j=1}^k \f(\xi_{j+1}, \xi_{j}) =
\prod_{i=1}^m \f(\omega_i, \omega_{i+1}) \cdot \prod_{j=1}^k \frac{1}{\f(\xi_j, \xi_{j+1})} = 1.
\] 
Therefore, \[ \prod_{i=1}^m \f(\omega_i, \omega_{i+1}) = \prod_{j=1}^k \f(\xi_j, \xi_{j+1}), \] shows that the definition of \(\f(\omega_1, \omega_{m+1})\) does not depend on the particular sequence and is therefore well defined. 
\hfill  \qed

\subsection{The proof of Proposition~\ref{Proposition: basic charecterization}.} \label{Appendix: proof of first main result}

\heading{Part 1: Necessity.} 
Suppose that there is an $F$-measurable positive function $f$ that satisfies Eq.~\eqref{eq: theorem 1}. 
To show {\rm{\textbf{[INC]}}}, let $\bigl((\o_i, \o_{i+1})\bigr)_{i=1}^m$ be an $F$-cycle. Then, 
 $$ 
\prod_{i=1}^m \f(\o_i,{\o_{i+1}})=\prod_{i=1}^m  \frac{f(\o_{i})}{f(\o_{i+1})}=
\frac{f(\o_{1})}{f(\o_{m+1})}=1.
 $$
The first equality is due to  Eq.~\ref{eq: theorem 1}, and the last follows from the assumption that $f$ is $F$-measurable and that 
 $\o_{m+1} \in F(\o_{1}).$
 
To show \textbf{[EXC]}, let \(\bigl((\omega_i, \bar{\omega}_i)\bigr)_{i=1}^m\) 
be an $F$-loop. 
By Eq.~\eqref{eq: External con} and Eq.~\eqref{eq: theorem 1}, we have
\[
\prod_{i=1}^m \f(\omega_i, \bar{\omega}_i)
= \prod_{i=1}^m \frac{f(\omega_i)}{f(\bar{\omega}_i)}
=\frac{f(\omega_1)}{f(\bar{\omega}_m)} = 1.
\]
The first equality follows from Eq.~\eqref{eq: theorem 1}, the second equality follows from the fact that $f$ is $F$-measurable and the third property of $F$-loops, and the third equality follows from the same property, since \(\bar{\omega}_m \in F(\omega_1)\).

\heading{Part 2: Sufficiency.} 
Assume \textbf{[INC]} and \textbf{[EXC]} with the extension provided by Lemma~\ref{lemma:extension}.
We claim first that for every connected set of $V$, say $C$, there is an $F$-measurable positive function $f_C$, defined over $C$, that satisfies  Eq.~\eqref{eq: theorem 1}. 
Let $\o_0$ be an arbitrary state in C. Set $f(\o_0)=1$. 
We proceed by induction on the distance\footnote{The distance between two states in 
$C$ is defined as the minimal number of edges in a path connecting them.} to $\o_0$. 

Suppose that \(f\) has been defined on all states in a connected component \(C\) whose distance from the state \(\omega_0\) is at most \(k\). 
Let \(\omega'\) be a state in \(C\) at distance \(k+1\) from \(\omega_0\). 
Then, there exists a state \(\omega''\) at distance \(k\) such that \((\omega'', \omega') \in E\). 
Define
\[
f_C(\omega') := f_C(\omega'') \cdot \f(\omega', \omega'').
\]
Using \textbf{[INC]} and the same argument as in Lemma~\ref{lemma:extension}, one can show that \(f\) is well defined. To show that $f_C$ is $F$-measurable, consider $\o \in F( \o') \cap C  $. Then, there is an F-cycle $\o_1=\o,...,\o_{m+1}=\o'$. 
By \textbf{[INC]} and by the definition of $f_C $,
\[1=
\prod_{i=1}^m \f(\omega_i, {\omega}_{i+1}) 
= \prod_{i=1}^m \frac{f_C(\omega_i)}{f_C({\omega}_{i+1})}
=\frac{f_C(\omega_1)}{f_C({\omega}_{m+1})}
=\frac{f_C(\omega)}{f_C({\omega}')}.
\]
Thus, the values that $f_C$ takes on 
$\o $ and $ \o' $ coincide.
So far, we have constructed \(f_C\) for each connected component \(C\) of \(V\). 
We now show that there exists a single $F$-measurable function \(f\) satisfying Eq.~\eqref{eq: theorem 1}. 

To this end, we define a new graph, $\bar{G}= (\bar{V},\bar{E})$ as follows. The set of vertices, $\bar{V}$, consists of the connected components of $V$. Two components $C, C' \in \bar{V}$ are connected by an edge if there exist states $\omega \in C$ and $\omega' \in C'$ such that $\omega \in F(\omega')$. That is, the edge set,  $\bar{E}$, consists of all pairs $(C, C')$ for which there exist $\omega \in C$ and $\omega' \in C'$ such that $\omega$ and $\omega'$ belong to the same information set of $F$.

Apply the following construction separately to each connected component of $\bar G$; no $F$-cell joins different components. Fix a vertex \(C_0\), and set \(f := f_{C_0}\) on \(C_0\). We proceed by induction on the distance of a vertex \(C\) from \(C_0\).  Suppose that an $F$-measurable function \(f>0\), satisfying Eq.~\eqref{eq: theorem 1}, has been defined on all vertices whose distance from \(C_0\) is less than or equal to \(k\). Let \(C\) be a vertex at distance \(k+1\) from \(C_0\). Then there exists a path of edges in \(\bar{E}\),
\[
(C_0, C_1), (C_1, C_2), \dots, (C_k, C),
\]
and in particular, there exist \(\bar{\omega}_k \in C_k\) and \(\omega \in C\) such that \(\bar \omega_k\) and \(\omega\) belong to the same information set of F. Define
\begin{equation} \label{eq:extension-f-all-Vbar}
    f := f_C \cdot \frac{f(\bar{\omega}_k)}{f_C(\omega)}
\end{equation}
on \(C\). Clearly, if \(f\) is well defined in \(C\), then it satisfies Eq.~\eqref{eq: theorem 1}, since it differs from \(f_C\) by a multiplicative factor of a positive constant. Moreover, 
$f(\omega) = f(\bar{\omega}_k)$.

We first verify that $f$ is well defined and then show that it is $F$-measurable. To establish that $f$ is well defined, suppose that there exists another shortest path 
\[
(C_0, C'_1), (C'_1, C'_2), \dots, (C'_{\ell}, C),
\] 
connecting $C_0$ and $C$, with \({\omega}'_{\ell} \in C'_{\ell}\) and \(\omega' \in C\) such that
\(\omega'_{\ell}\) and \(\omega'\) belong to the same information set of F.\footnote{The asymmetry in the notation between $\bar{\omega}_k$ (with a bar) and  ${\omega}'_{\ell}$ (without) will become clear immediately.} 
We have to show that
\begin{equation} \label{eq: f well defined}
\frac{f(\bar{\omega}_k)}{f_C(\omega)}= \frac{f({\omega}'_{\ell})}{f_C(\omega')}.    
\end{equation}

If $k=0$, then $\ell=0$ and $C'_\ell=C_0$. Applying \textbf{[EXC]} to the $F$-loop $((\omega'_\ell, \bar\omega_k), (\omega,\omega'))$ gives $\tfrac{f(\omega'_\ell)}{f(\bar\omega_k)}\cdot \tfrac{f_C(\omega)}{f_C(\omega')}=1$, which proves Eq.~\eqref{eq: f well defined}. 

Otherwise, note that the path 
\[
(C_0, C_1), (C_1, C_2), \dots, (C_k, C), (  C, C'_{\ell}), \dots,(C'_2, C'_1), (C'_1, C_0)
\]
is a cycle in the graph $\bar{G}$,  that corresponds to an F-loop:
\[
(\o_0,\bar{\o}_0), (\o_1, \bar{\o}_1), \dots, (\o_{k-1}, \bar{\o}_{k-1}),(\o_k, \bar{\o}_k), 
(\o , {\o'}), (\o'_{\ell}, \bar{\o}'_{\ell}), \dots,(\o'_2,  \bar{\o}'_2)
, (\o'_1, \bar{\o}'_1). 
\]
That is, any two states in a pair belong to the same vertex in $\bar{G}$, e.g., $\o_0$ and $\bar{\o}_0$ belong to $C_0$, $\o_1$ and $\bar{\o}_1$ belong to $C_1$; $\o$ and ${\o'}$ belong to $C$, and $\o'_{j}, \bar{\o}'_{j}$ belong to $C'_j$.   Also, in two adjacent pairs, the second coordinate of the left pair belongs to the same information set of F as the first coordinate of the right pair.  For instance, $\bar{\o}_i\in F(\o_{i+1})$, $\bar{\o}_k \in F(\o)$, ${\o'} \in F(\o'_{\ell})$ and ${\bar \o}'_{j+1}\in F({\o}'_j)$.  Finally, $\bar{\o}'_1\in F(\o_0)$. 

By  \textbf{[EXC]}, and specifically by  Eq.~\eqref{eq: ECON}, the corresponding product equals 1.  That is,
\begin{equation}\label{eq: product of loop}
   \left[\prod_{i=0}^k \f(\o_i, \bar{\o}_i) \right] \cdot \f(\o ,{\o'}) \cdot \left[ \prod_{j=1} ^{\ell} \f (\o'_{j}, \bar{\o}'_{j}) \right] =1. 
\end{equation}

Due to F-measurability and to Eq.~\eqref{eq: theorem 1},  we get $\prod_{i=0}^k \f(\o_i, \bar{\o}_i)=\frac{f(\o_0)}{f(\bar{\o}_k)}$, $\f(\o ,{\o'})= \frac{f_C(\o)}{f_C(\o')} $, and $\prod_{j=1}^{\ell} \f (\o'_{j}, \bar{\o}'_{j})= \frac{f(\o'_{\ell})}{f(\bar{\o}'_1)}$.  Thus, by Eq.~\eqref{eq: product of loop} we obtain the following:
\[
\frac{f(\o_0)}{f(\bar{\o}_k)}\cdot  \frac{f_C(\o)}{f_C(\o')} \cdot\frac{f(\o'_{\ell})}{f(\bar{\o}'_1)} =1. 
\]
Due to F-measurability $f(\o_0)= f(\bar{\o}'_1)$ and we conclude  that 
\[
\frac{f(\o'_{\ell})} {f(\bar{\o}_k)}\cdot  \frac{f_C(\o)}{f_C(\o')}  =1. 
\]
which confirms Eq.~\eqref{eq: f well defined}.

The final step in the proof is to show that \(f\) is $F$-measurable. Within each component, $F$-measurability follows from that of $f_C$. For distinct components $C$ and $C'$, we consider two cases:

\paragraph{Case I.} Let \(\omega \in C\) and \(\omega' \in C'\), where the distance from \(C'\) to \(C_0\) is less than or equal to \(k\), and suppose that \(\omega \in F(\omega')\). We need to show that \(f(\omega) = f(\omega')\). By assumption, \((C, C') \in \bar{E}\) and \((C, C_k) \in \bar{E}\). Moreover, \(C'\) and \(C_k\) are connected to $C_0$ by paths of length at most \(k\). This forms a cycle in the graph \(\bar{G}\), which in turn corresponds to an \(F\)-loop. Applying the same technique used earlier, we conclude that \(f(\omega) = f(\omega')\).

\paragraph{Case II.} Let \(\omega \in C\) and \(\omega' \in C'\), where the distance from \(C'\) to \(C_0\) is exactly \(k+1\), so that both \(C\) and \(C'\) are introduced in the induction process at the same step, and assume that \(\omega \in F(\omega')\). Then \((C, C') \in \bar{E}\), and both \(C\) and \(C'\) are connected to \(C_0\) by paths of length \(k+1\). This forms a cycle in \(\bar{G}\) passing through \(C\) and \(C'\), which corresponds to an \(F\)-loop. Using \textbf{[EXC]}, we conclude that \(f(\omega) = f(\omega')\), verifying that \(f\) is \(F\)-measurable.
 \hfill \qed

\subsection{The proof of Theorem \ref{th: main Bayesian}} \label{Section  - Proof of second theorem}

\begin{proof}
Assume that $\tau$ is a signaling function and that $s$ is a signal generated by $\tau$ with positive probability such that $\boldsymbol{\mu}_{\tau,s} = \text{JB}$. 
Then, clearly, the set of states $\omega$ such that $\tau(s \mid \omega) > 0$ is measurable w.r.t.\ the mediator's partition $F$, and matches the set $\Omega_+$, previously defined according to JB.

Similarly to Eq.~\eqref{eq:likelihood}, the equality $\boldsymbol{\mu}_{\tau,s} = \text{JB}$ implies that $\tfrac{\tau(s|\o)\mu(\o)}{\tau(s|\o')\mu(\o')} = \tfrac{{\rm{JB}}(\omega,i)(\o)}{{\rm{JB}}(\omega',i)(\o')}$.
Applying Eq.~\eqref{eq: defining phi}, we get
$$
\f_{\text{JB}}(\omega, \omega') = \frac{{\rm{JB}}(\omega,i)(\o)}{{\rm{JB}}(\omega',i)(\o')} \cdot \frac{\mu(\omega')}{\mu(\omega)} = \frac{\tau(s|\o)}{\tau(s|\o')},
$$
and Proposition~\ref{Proposition: basic charecterization} guarantees that $\f_{\text{JB}}$ satisfies conditions {\rm\textbf{[INC]}} and {\rm\textbf{[EXC]}}  in $G_+$. 
This proves the necessary direction.

For the sufficiency direction, assume that JB satisfies conditions (i) and (ii).  Since $\Omega_+$ is measurable w.r.t.\ $F$, we can define $\tau(s \mid \omega) = 0$ for every $\omega \notin \Omega_+$, and otherwise, $\tau(s \mid \omega)$ is positive. 

On $\Omega_+$, since $\f_{\text{JB}}$ satisfies {\rm\textbf{[INC]}} and {\rm\textbf{[EXC]}} in $G_+$, Proposition~\ref{Proposition: basic charecterization} ensures the existence of a function $f > 0$ satisfying Eq.~\eqref{eq: theorem 1} and we get
$$
\frac{{\rm{JB}}(\omega,i)(\o)}{{\rm{JB}}(\omega',i)(\o')} \cdot \frac{\mu(\omega')}{\mu(\omega)} = \frac{f(\o)}{f(\o')},
$$
By multiplying $f$ by a constant if needed, we may assume without loss of generality that $\sup_{\o} f(\o) < 1$. We then define $\tau(s \mid \omega) = f(\omega)$ for all $\omega \in \Omega_+$, so that the previous equation can be written as 
$$
\frac{\tau(s \mid \omega) \mu(\o)}{\tau(s \mid \omega') \mu(\o')} = \frac{{\rm{JB}}(\omega,i)(\o)}{{\rm{JB}}(\omega',i)(\o')},
$$
which matches Eq.~\eqref{eq:likelihood}. By condition (ii) and information-set invariance, both distributions have support $\mathcal{P}_i(\omega)\cap\Omega_+$; matching their probability ratios therefore implies equality. This guarantees that $\tau(s \mid \cdot)$ implements JB, as needed.  To ensure that $\tau$ is a well-defined kernel, choose a signal $s_0\neq s$, set $S=\{s,s_0\}$, and define $\tau(s_0\mid\omega)=1-\tau(s\mid\omega)$ for every $\omega\in\Omega$. In particular, $\tau(s_0\mid\omega)=1$ for every $\o \in \Omega\setminus \Omega_+$. Both signal likelihoods are $F$-measurable, and their probabilities sum to one at every state.
\hfill
\end{proof}

\subsection{The proof of Theorem \ref{Th:convexity}} \label{Section - proof of convexity theorm}

\begin{proof}
We first prove that no fully informative $F$-loops implies pseudo-independence. Define the bipartite graph $G=(\mathcal{P}_1,F, E)$ where $(p,f)\in E$ if there exists $\omega \in p\cap f$. So two vertices $p$ and $f$ are connected if and only if they share a state $\omega\in \Omega$. A cycle exists in $G$ if and only if a fully informative $F$-loop exists. Indeed, selecting one state from each intersection along a simple cycle
produces a fully informative $F$-loop. Conversely, a fully informative $F$-loop determines a closed walk in $G$
that never immediately retraces an edge: successive CKCs are distinct by the loop definition, and the incoming and outgoing $F$-cells at each CKC are distinct by full informativeness. Such a closed walk contains a simple cycle. We therefore conclude that each connected component in $G$ is a tree.

For every $p\in \mathcal{P}_1$ and $f\in F$ define $m_{p,f}=\mu(p\cap f)$. Take a tree in $G$, denote a root by $p_0 \in \mathcal{P}_1$ and define $a_{p_0}=1$. For every $f_0$ connected to $p_0$ in $G$ define $b_{f_0}=\mu(p_0\cap f_0)$. Fix an $f_0$ and for every $p_1$ connected to it (in $G$; not including $p_0$) define $a_{p_1}=\tfrac{\mu(p_1\cap f_0)}{b_{f_0}}$. Continue inductively and similarly for every tree in $G$, until all parameters $a_p$ and $b_f$ are defined, to conclude that $\mathcal{P}_1$ and $F$ are pseudo-independent.

Next, we prove that the stated pseudo-independence implies a convex feasible set.  Consider two feasible profiles under $\mu$. First suppose both have full support. For each $p\in\mathcal{P}_1$,  let $q_p$ be the convex combination of their posteriors in $p$, using the same weight $\lambda\in[0,1]$ for every $p$.  Each original posterior, and hence $q_p$, preserves the prior proportions within every nonempty intersection $p\cap f$. Define a strictly positive probability distribution $\nu$ by $\nu(\omega)=\mu(p)q_p(\omega)$ for every $\omega\in p$. Thus, $\nu(\cdot\mid p)=q_p$ and $\nu(\cdot\mid p\cap f)=\mu(\cdot\mid p\cap f)$. Applying pseudo-independence to both $\mu$ and $\nu$, choose positive factors (taking absolute values if necessary) such that $\mu(p\cap f)=a_pb_f$ and $\nu(p\cap f)=c_pd_f$ whenever $p\cap f\neq\emptyset$.

For $\omega\in p\cap f$, it follows that
\[
\frac{\nu(\omega)}{\mu(\omega)} =\frac{\nu(p\cap f)}{\mu(p\cap f)} =\frac{c_p}{a_p}\cdot \frac{d_f}{b_f}.
\]
Choose $\kappa>0$ sufficiently small that $\kappa d_f/b_f<1$ for every $f\in F$, and define $\tau(s\mid\omega)=\kappa\tfrac{d_{F(\omega)}}{b_{F(\omega)}}$ and $\tau(s_0\mid\omega)=1-\tau(s\mid\omega)$. This is an $F$-measurable experiment. For $\omega\in p$, we get $\mu(\omega)\tau(s\mid\omega)=\kappa\frac{a_p}{c_p}\nu(\omega)$. The multiplier is constant $\kappa\frac{a_p}{c_p}$ within $p$, so Bayes' rule gives $\mu_{\tau,1}(\cdot\mid p,s)=\nu(\cdot\mid p)=q_p$. Hence the feasible set is convex.

For profiles with the same smaller support $\Omega_+$, restrict both partitions and the prior to this $F$-measurable set. Pseudo-independence is inherited by the restricted partitions: any positive prior on $\Omega_+$ can be obtained by conditioning a positive prior on $\Omega$. Retain its factors on the nonempty restricted cells and divide the $F$-factors by the mass of $\Omega_+$. Apply the preceding construction on $\Omega_+$, set $\tau(s\mid\omega)=0$ outside $\Omega_+$, and use the complementary signal $s_0$. This proves convexity for profiles with matching support.

For the last part, suppose that the feasible set is convex for every prior but a fully informative $F$-loop exists. Choose a shortest cycle in $G$ and let $L$ be the fully informative loop obtained by selecting one state from each intersection along it. Its CKCs and connecting $F$-cells are respectively pairwise distinct, and $\overline{\o}_1\notin F(\o_1)$. For every $\o\notin L$, define $\mu(\o)=0$ (we can later assign a sufficiently small probability for any such state to maintain a strictly positive prior), and for every $\o\in L\setminus \{\overline{\o}_1\}$ set $\mu(\overline{\o}_1)=2\mu(\o)$. This establishes a prior $\mu$.

Define the signaling function $\tau$ as follows: $\tau(s_1|F(\overline{\o}_1))=1/3=1-\tau(s_2|F(\overline{\o}_1))$ and  $\tau(s_1|F \setminus F(\overline{\o}_1))=2/3=1-\tau(s_2|F \setminus F(\overline{\o}_1))$. This strategy induces two feasible posterior beliefs, one for each signal, defined over the loop $L$.  We can take the average of two posterior beliefs and check whether it is also feasible, specifically using the external consistency condition across the loop.

Notice that in every partition element $p \neq  \mathcal{P}_1(\o_1),\mathcal{P}_1(\o_2)$, we have a uniform prior and non-informative signals, so the posteriors match the priors and the likelihood ratios for external consistency, as in Eq.~\ref{eq: defining phi}, equal $1$. This ratio holds for the partition element $\mathcal{P}_1(\o_2)$ as well, because the prior is uniform, so
$$
\frac{\tfrac{1}{2}\mu_{\tau,1}(\o_2 \mid \mathcal{P}_1(\o_2),s_1) + \tfrac{1}{2}\mu_{\tau,1}(\o_2 \mid \mathcal{P}_1(\o_2),s_2) }{\tfrac{1}{2}\mu_{\tau,1}(\overline{\o}_2 \mid \mathcal{P}_1(\overline{\o}_2),s_1) + \tfrac{1}{2}\mu_{\tau,1}(\overline{\o}_2 \mid \mathcal{P}_1(\overline{\o}_2),s_2)}\cdot \frac{\mu(\overline{\o}_2)}{\mu({\o}_2)} = \frac{\tfrac{1}{3}+\tfrac{2}{3}}{\tfrac{2}{3}+\tfrac{1}{3}}\cdot 1 = 1. 
$$
Hence, the external consistency condition reduces to the likelihood ratio induced in $\mathcal{P}_1(\o_1)$, however 
$$
\frac{\tfrac{1}{2}\mu_{\tau,1}(\o_1 \mid \mathcal{P}_1(\o_1),s_1) + \tfrac{1}{2}\mu_{\tau,1}(\o_1 \mid \mathcal{P}_1(\o_1),s_2) }{\tfrac{1}{2}\mu_{\tau,1}(\overline{\o}_1 \mid \mathcal{P}_1(\overline{\o}_1),s_1) + \tfrac{1}{2}\mu_{\tau,1}(\overline{\o}_1 \mid \mathcal{P}_1(\overline{\o}_1),s_2)}\cdot \frac{\mu(\overline{\o}_1)}{\mu({\o}_1)}= \frac{\tfrac{1}{2}+\tfrac{1}{5}}{\tfrac{1}{2}+\tfrac{4}{5}} \cdot \frac{2\mu(\o_1)}{\mu(\o_1)}= \frac{14}{13}\neq 1.
$$
We thus conclude that external consistency fails and the average of the two posterior beliefs is not feasible, implying that the feasible set in not convex. 
\hfill
\end{proof}

\subsection{The Proof of Proposition \ref{prop:generic-nonconvex-informative-loop} } \label{Appendix: proof of generic non-convexity}

\begin{proof}
By the loop--cycle equivalence established in the proof of
Theorem~\ref{Th:convexity}, the fully informative $F$-loop hypothesis
implies that the bipartite incidence graph contains a cycle.
Choose a shortest cycle and select one state from each intersection
along it. Denote the resulting fully informative loop by
$\bigl((\o_i,\bar{\o}_i)\bigr)_{i=1}^{m}$.
Write $p_i=\mathcal{P}_1(\o_i)$ and $f_i=F(\bar{\o}_i)$.
The $p_i$s are pairwise distinct, as are the $f_i$s.
Since the cycle is shortest, $p_i\cap f_1=\emptyset$ for $i\geq3$.

Fix two public signals $s_1,s_2$ and define an $F$-measurable kernel $\tau$ by $\tau(s_1\mid f_1)=\frac13$, $\tau(s_1\mid f)=\frac23$ for every $f\in F\setminus\{f_1\}$, and $\tau(s_2\mid f)=1-\tau(s_1\mid f)$ for every $f\in F$.
Since $\mu$ has full support, both $s_1$ and $s_2$ occur with positive probability at every state.
Let ${\rm{JB}}^{(1)}$ and ${\rm{JB}}^{(2)}$ denote the joint beliefs induced by conditioning on $s_1$ and $s_2$, respectively. 
Both are feasible by construction and have matching support (given any state), so their average ${\rm{JB}}^{(\mathrm{avg})}:=\frac12\,{\rm{JB}}^{(1)}+\frac12\,{\rm{JB}}^{(2)}$ is a well-defined joint belief.

To maintain convexity, \textbf{[EXC]} must hold given $\f_{{\rm{JB}}^{(\mathrm{avg})}}$.
Because $s_1$ and $s_2$ are completely uninformative in every $p_i\neq p_1,p_2$ along the loop, $\f_{{\rm{JB}}^{(\mathrm{avg})}}(\o_i,\bar{\o}_i)=1$ for every $i=3,4,\dots,m$.
Thus,  \textbf{[EXC]} reduces to $\prod_{i=1}^2 \f_{{\rm{JB}}^{(\mathrm{avg})}}(\o_i,\bar{\o}_i)=1$.

For any CKC $p \in \{p_1,p_2\}$ define the (prior) mass of the mediator partition element $f_1$ inside $p$ by $x_p := \tfrac{\mu(p\cap f_1)}{\mu(p)}\in(0,1)$, and let $D_p^{(k)} = \sum_{f\in F}\tau(s_k\mid f)\mu(p\cap f)$ be the (unnormalized) probability that signal $s_k$ occurs given $p$.
We thus get
\[
D_p^{(1)}=\frac13\,\mu(p\cap f_1)+\frac23\big(\mu(p)-\mu(p\cap f_1)\big)
=\frac{\mu(p)}{3}(2-x_p),
\]
\[
D_p^{(2)}=\frac23\,\mu(p\cap f_1)+\frac13\big(\mu(p)-\mu(p\cap f_1)\big)
=\frac{\mu(p)}{3}(1+x_p).
\]
Because ${\rm{JB}}^{(\mathrm{avg})}$ is an average of two joint beliefs, for any $\omega\in p\cap f$, we get
\[
{\rm{JB}}^{(\mathrm{avg})}(\omega,1)(\omega)
= \mu(\omega)\left(\frac12\cdot \frac{\tau(s_1\mid f)}{D_p^{(1)}}+\frac12\cdot \frac{\tau(s_2\mid f)}{D_p^{(2)}}\right).
\]
We can now substitute these values in the simplified \textbf{[EXC]} condition above, and get
\begin{eqnarray*}
 1 
 & = &  \frac{\mu(\o_1)\cdot \left(\tfrac12\cdot \tfrac{2/3}{D_{p_1}^{(1)}} +\tfrac12\cdot \tfrac{1/3}{D_{p_1}^{(2)}} \right)}{ \mu(\bar{\o}_1)\cdot \left(\tfrac12\cdot \tfrac{1/3}{D_{p_1}^{(1)}} +\tfrac12\cdot \tfrac{2/3}{D_{p_1}^{(2)}} \right)   } \cdot \frac{\mu(\bar{\o}_1)}{\mu(\o_1)} \cdot \frac{\mu(\o_2)\cdot \left(\tfrac12\cdot \tfrac{1/3}{D_{p_2}^{(1)}} +\tfrac12\cdot \tfrac{2/3}{D_{p_2}^{(2)}} \right)}{ \mu(\bar{\o}_2)\cdot \left(\tfrac12\cdot \tfrac{2/3}{D_{p_2}^{(1)}} +\tfrac12\cdot \tfrac{1/3}{D_{p_2}^{(2)}} \right)   } \cdot \frac{\mu(\bar{\o}_2)}{\mu(\o_2)} \\
 & = & \frac{ \left( \tfrac{2}{D_{p_1}^{(1)}} + \tfrac{1}{D_{p_1}^{(2)}} \right)}{  \left(\tfrac{1}{D_{p_1}^{(1)}} + \tfrac{2}{D_{p_1}^{(2)}} \right)   } \cdot \frac{\left( \tfrac{1}{D_{p_2}^{(1)}} + \tfrac{2}{D_{p_2}^{(2)}} \right)}{ \left( \tfrac{2}{D_{p_2}^{(1)}} + \tfrac{1}{D_{p_2}^{(2)}} \right)   }   =  \frac{ \left( \tfrac{2}{2-x_{p_1}} + \tfrac{1}{1+x_{p_1}} \right)}{  \left(\tfrac{1}{2-x_{p_1}} + \tfrac{2}{1+x_{p_1}} \right)   } \cdot \frac{\left( \tfrac{1}{2-x_{p_2}} + \tfrac{2}{1+x_{p_2}} \right)}{ \left( \tfrac{2}{2-x_{p_2}} + \tfrac{1}{1+x_{p_2}} \right) } \\
 & = & \frac{4+x_{p_1}}{5-x_{p_1}}\cdot  \frac{5-x_{p_2}}{4+x_{p_2}},
\end{eqnarray*}
where the third equality follows from substituting the different values of $D_{p}^{(i)}$.

Define the function $R(x)=\frac{4+x}{5-x}$ and note it is strictly increasing on $(0,1)$.
Thus, the last equality is equivalent to 
\begin{equation}\label{eq:knife-edge}
R(x_{p_1})=R(x_{p_2}) \  \Leftrightarrow \  x_{p_1} = x_{p_2} \ \Leftrightarrow \mu(p_1\cap f_1) \mu(p_2) - \mu(p_2\cap f_1) \mu(p_1)=0.
\end{equation}
Let $\mathcal{N}\subset\Delta(\Omega)$ be the set of priors satisfying Equation~\eqref{eq:knife-edge}.
The left-hand side of Equation~\eqref{eq:knife-edge} is a nonzero polynomial in the coordinates of $\mu$, hence its zero set $\mathcal{N}$ has Lebesgue measure zero in the interior simplex. 
Therefore, for every full-support prior, excluding a set of Lebesgue measure zero, the feasible set is not convex.
\end{proof}

\subsection{The proof of Theorem \ref{Theorem: multiple JP}} \label{Appendix: proof of third theorem}

\begin{proof}
\heading{Part (i)}
We start by proving necessity. Suppose there exists an \({F}\)-measurable signaling function \(\tau\) as described in the theorem.  The induced posteriors thus form a martingale.  That is, if the posterior \(\mu_{i}\) is realized with probability \(q_i\), where \(\sum_i q_i = 1\), then
\[
\sum_i q_i \mu_{i}= \mu.
\]
Let \(u\) be an option. So,
\[
\E_{\mu}(u) = \sum_i q_i \E_{\mu_{i}}(u).
\]
If each \(\E_{\mu_{i}}(u) \ge 0\), then \(\E_{\mu}(u) \ge 0\) as well. Therefore, the family \(\{\mu_{1}, \ldots, \mu_m\}\) PP w.r.t.\ \(\mu\).

Moving on to prove sufficiency,   suppose that no \({F}\)-measurable signaling function \(\tau\) exists such that all the posteriors it generates are in \(\{\mu{_1}, \ldots, \mu{_m}\}\).  Note however, that every $\mu_i$ can be generated through a specific $F$-measurable strategy and appropriate signal, and also note that convex combinations of $F$‑measurable kernels are $F$‑measurable as well.  Thus, by standard results on Blackwell experiments, if $\mu\in\operatorname{conv}\{\mu_1,\ldots,\mu_m\}$, then there is an $F$-measurable experiment whose posteriors are contained in $\{\mu_1,\ldots,\mu_m\}$.

Our assumption implies that \(\mu \notin \operatorname{conv}\{\mu_{1}, \ldots, \mu_m\}\). Since this convex hull is closed, the separating hyperplane theorem guarantees the existence of a nonzero vector \(u \in \mathbb{R}^{\Omega}\) and a constant $c$, such that
\[
\E_{\mu_{i}}(u) = \langle \mu_{i}, u \rangle > c \quad \text{for all } i,
\]
while
\[
\E_{\mu}(u) = \langle \mu, u \rangle < c.
\]
By subtracting $c$ in the two inequalities above and replacing \(u\) with \(u - (c, \ldots, c)\), noting that the separation theorem is applied here to probability distributions, we conclude that the family \(\{\mu_{1}, \ldots, \mu_m\}\) does not PP w.r.t.\  \(\mu\).
This is a contradiction.

\heading{Part (ii)} We prove necessity first. Suppose there exists an \({F}\)-measurable signaling function \(\tau\) as described in Part (ii) of the theorem. Let $u$ be an option  such that $\E_{\mu}(u) = 0$ and $\E_{\mu_i}(u) \ge 0$ for every $i$.  We show that $\E_{\mu_i}(u) = 0$ for every $i$.  As before, if the posterior \(\mu_{i}\) is realized with probability \(q_i >0\), where \(\sum_i q_i = 1\), then
\[
\sum_i  q_i \mu_{i} = \mu.
\]
Let \(u\) be an option. So,
\[
\E_{\mu}(u) = \sum_{i=1}^m q_i \E_{\mu_{i}}(u).
\]
If each \(\E_{\mu_{i}}(u) \ge 0\) and $\E_{\mu}(u) = 0$, then \(\E_{\mu_{i}}(u) = 0\) for every $i$ . Therefore, the family \(\{\mu_{1}, \ldots, \mu_m\}\) SPP w.r.t.\ \(\mu\).

To establish sufficiency and similarly to the proof of Part (i), we show that $\mu$ can be expressed as a convex combination of $\mu_{1}, \ldots, \mu_m$, each with a positive weight. 
For this purpose, we prove the two following lemmas.

\begin{lemma}\label{lemma 2}
\emph{SPP} implies that for every $j$ there is a convex combination $\mu=\sum_i  q_i \mu_{i}$, where $q_j>0$.
\end{lemma}

\proof
 Fix an index~$j$. Suppose, to the contrary, that there is no convex combination  \(\mu = \sum_i q_i \mu_{i}\) with~$q_j > 0$.  This implies that the set
\[
D :=  \Bigl\{ q_j \mu_{j} + \sum_{i \ne j} q_i \mu_{i} \; ; \; q_j > 0,\; q_i \ge 0~\text{for}~i \ne j,~\text{and}~q_j + \sum_{i \ne j} q_i = 1 \Bigr\}
\]
does not contain~$\mu$.  Note that~$D$ is convex and has nonempty relative interior in $\operatorname{aff}\{\mu_1,\ldots,\mu_m\}$; in particular, $\frac{1}{m}\sum_i\mu_i$ is a relative interior point of~$D$.

Consider now the closure of~$D$, denoted~$\bar D$.  This set is also convex and contains all the~$\mu_{i}$’s (including~$\mu_{j}$).  The point~$\mu$ may lie outside~$\bar D$ or on its boundary.  By Theorem~11.3 in~\citealp{rockafellar1970convex}, proper separation of $D$ and $\{\mu\}$ gives a nonzero option~$u$ such that \(\E_\mu(u)=0\) and \(\E_\nu(u)\ge0\) for every \(\nu\in\bar D\), with strict inequality for some \(\nu\in\bar D\).\footnote{For a separating option $v$, take $u=v-\E_\mu(v)\mathbf{1}$.} In particular, this implies that  \(\E_{\mu_{i}}(u) \ge 0\)  for every~$i$.

We now apply SPP to the option~$u$, which yields  \(\E_{\mu_{i}}[u] = 0\)  for all~$i$. Consider, however, the relative interior point  \(\nu = \frac{1}{m}\sum_i \mu_{i}\),  which belongs to~$\bar D$.  By proper separation and relative interior, we have on the one hand \(\E_{\nu}[u] > 0\), while on the other hand,
\[
\E_{\nu}[u]  = \E_{\frac{1}{m}\sum_i \mu_{i}}[u]  = \frac{1}{m}\sum_i \E_{\mu_{i}}[u] = 0.
\]
This is a contradiction.  Hence, there must exist a convex combination  \(\mu = \sum_i q_i \mu_{i}\) with~$q_j > 0$, as required.
\hfill \qed

Given Lemma~\ref{lemma 2}, we now turn to the next lemma, which shows that $\mu$ can be expressed as a convex combination of $\mu_{1}, \ldots, \mu_m$, each with a positive weight.

\begin{lemma}\label{lemma 1} 
If for every $j$ there is a convex combination of $\mu_{1}, \ldots, \mu_m$ such that $\mu=\sum_i q_i \mu_{i} $, where $q_j>0$, then there exists a combination where all $q_i$ are strictly positive. 
\end{lemma}
\proof Fix $j$ and suppose that $\mu=\sum_i q_i^j  \mu_{i} $ is  a convex combination, where $q_j^j>0$. Take $\mu=\frac{1}{m}\sum_{j=1}^m\sum_i q_i^j\mu_{i} $, and note that in this combination, the weight of every $\mu_{i}$ is positive.
\hfill \qed 

Combining the two lemmas, we derive from SPP that $\mu$ is a convex combination (with strictly positive weights) of $\mu_{1}, \ldots, \mu_m$. Applying the construction in Part (i) with these positive weights completes the proof.
\hfill 
\end{proof}

\end{document}